\documentclass{jfm}
\usepackage{natbib}
\usepackage{definitions}
\usepackage{localdefs}
\usepackage{amsmath}
\usepackage{amssymb}
\usepackage{graphics}
\usepackage{amsbsy}
\usepackage{upmath}

\title[Spherical particles between planar walls] 
{Hydrodynamic interactions of spherical particles in suspensions
confined between two planar walls.}

\author[S. Bhattacharya, J. B{\l}awzdziewicz and E. Wajnryb ]
{S.\ns B\ls H\ls A\ls T\ls T\ls A\ls C\ls H\ls A\ls R\ls Y\ls A,\break
J.\ns B\ls \L\ls A\ls W\ls Z\ls D\ls Z\ls I\ls E\ls W\ls I\ls C\ls Z,
\ns\and E.\ns W\ls A\ls J\ls N\ls R\ls Y\ls B%
\thanks{On leave from IPPT Warsaw, Poland}
\affiliation{Department of Mechanical Engineering, Yale University, New
Haven, CT 06520-8286, USA}
}

\date{}

\begin{document}
\maketitle

\begin{abstract}

Hydrodynamic interactions in a suspension of spherical particles
confined between two parallel planar walls are studied under
creeping-flow conditions.  The many-particle friction matrix in this
system is evaluated using our novel numerical algorithm based on
transformations between Cartesian and spherical representations of
Stokes flow.  The Cartesian representation is used to describe the
interaction of the fluid with the walls and the spherical
representation is used to describe the interaction with the particles.
The transformations between these two representations are given in a
closed form, which allows us to evaluate the coefficients in linear
equations for the induced-force multipoles on particle surfaces.  The
friction matrix is obtained from these equations, supplemented with
the superposition lubrication corrections.  We have used our algorithm
to evaluate the friction matrix for a single sphere, a pair of
spheres, and for linear chains of spheres.  The friction matrix
exhibits a crossover from a quasi-two-dimensional behavior (for
systems with small wall separation $H$) to the three-dimensional
behavior (when the distance $H$ is much larger than the interparticle
distance $L$).  The crossover is especially pronounced for a long
chain moving in the direction normal to its orientation and parallel
to the walls.  In this configuration, a large pressure buildup occurs
in front of the chain for small values of the gapwidth $H$, which
results in a large hydrodynamic friction force.  A standard wall
superposition approximation does not capture this behavior.

\end{abstract}

\section{Introduction}
\label{Introduction}

Numerous recent papers reflect a growing interest in the static and
dynamic properties of suspensions in confined geometries.  There are
investigations of the formation of colloidal crystals on patterned and
planar surfaces
\cite[][]{Lin-Crocker-Prasad-Schofield-Weitz-Lubensky-Yodh:2000,%
Seelig-Tang-Yamilov-Cao-Chang:2002,%
Subramanian-Manoharan-Thorne-Pine:1999}, 
studies of single-file diffusion of Brownian particles in a channel
\cite[][]{Wei-Bechinger-Leiderer:2000}, and experiments on
quasi-two-dimensional suspensions confined between two planar walls
\cite[][]{%
Carbajal_Tinoco-Cruz_de_Leon-Arauz_Lara:1997,%
Lancon-Batrouni-Lobry-Ostrowsky:2001,%
Marcus-Schofield-Rice:1999,%
Santana_Solano-Arauz_Lara:2001%
}.
Quasi-two-dimensional suspensions  of particles adsorbed 
at a fluid interface
\cite[][]{%
Zahn-Mendez_Alcaraz-Maret:1997,%
Rinn-Zahn-Maass-Maret:1999,
Cichocki-Ekiel_Jezewska-Nagele-Wajnryb:2004%
}
or confined in a thin liquid film
\cite*[][]{Sethumadhavan-Nikolov-Wasan:2001} have also been examined.

Experiments on quasi-two-dimensional systems revealed many striking
phenomena like, for instance, the first-order transitions between
fluid, hexatic, and solid phases \cite[][]{Marcus-Rice:1997},
string-like cooperative motion of suspension particles
\cite*[][]{Marcus-Schofield-Rice:1999}, and oscillatory melting of a
crystalline phase in shear flow
\cite[][]{Stancik-Hawkinson:2004}. Other interesting examples include
a logarithmic behavior of the mean-square displacement of Brownian
particles in quasi-two-dimensional systems, predicted by
\cite{Cichocki-Felderhof:1994e} and observed by
\cite{Marcus-Schofield-Rice:1999}; a hydrodynamic enhancement of
self-diffusion for strongly-charged particles
\cite[][]{Zahn-Mendez_Alcaraz-Maret:1997,Pesche-Nagele:2000}, and
migration of particles in Poiseuille flow towards the channel center
\cite[][]{Nott-Brady:1994,Lyon-Leal:1998a}.

The particle--wall and interparticle interaction potentials fully
determine the equilibrium structure of confined colloidal suspensions.
The dynamics of such systems, however, is significantly affected by
the many-body hydrodynamic forces.  For spherical particles in an
unbounded space, efficient algorithms for evaluating many-body
friction and mobility matrices have been developed
\cite*[][]{
Durlofsky-Brady-Bossis:1987,%
Ladd:1988,%
Cichocki-Felderhof-Hinsen-Wajnryb-Blawzdziewicz:1994,%
Sierou-Brady:2001%
}.
Using the image representation technique, such algorithms have been
generalized for particles adsorbed on a planar fluid--air interface
\cite[][]{Cichocki-Ekiel_Jezewska-Nagele-Wajnryb:2004} and for
particles confined in a thin liquid film
\cite[][]{Blawzdziewicz-Wajnryb:2003-C1}.  The image-representation
method has also been proposed for a suspension bounded by a single
rigid planar wall
\cite[][]{Cichocki-Jones:1998,Cichocki-Jones-Kutteh-Wajnryb:2000}.  A
two-wall generalization of this method
\cite[][]{Bhattacharya-Blawzdziewicz:2002} and several other
techniques were used to describe motion of an individual particle
between two planar walls
\cite[][]{Ganatos-Weinbaum-Pfeffer:1980,%
Ganatos-Pfeffer-Weinbaum:1980,%
Staben-Zinchenko-Davis:2003,%
Jones:2004%
}.  

The hydrodynamics of many particles in the two-wall geometry is much
more complex, and available results are limited.
\cite{Durlofsky-Brady:1989} have developed a method that combines
boundary-integral and Stokesian-dynamics elements.  In their approach,
the walls are discretized, and the particles are represented using
low-order force multipoles and lubrication contributions, as in the
standard Stokesian-dynamics algorithm
\cite[][]{Durlofsky-Brady-Bossis:1987}.  It seems that this method has
not been further pursued.  In an alternative approach,
\cite{Nott-Brady:1994} and \cite{Morris-Brady:1998} studied flows in
wall-bounded suspensions by modeling the walls as static, closely
packed arrays of spheres, and using the standard Stokesian-dynamics
algorithm for an unbounded system to evaluate the motion of the
suspended particles. The results obtained in this way are only
qualitative, especially for small wall separations, because the walls
are porous and rough.  Recently, a two-wall superposition
approximation was used by \cite{Pesche-Nagele:2000} and several other
groups, but the validity range of this approximation cannot be
determined without comparison with more accurate results.

In our paper we present a novel, highly accurate algorithm to evaluate
the many-body hydrodynamic interactions in a suspension of spherical
particles confined between two planar walls.  In our approach, the
flow field in the system is expanded using two basis sets of solutions
of Stokes equations---the spherical and Cartesian bases.  The
spherical basis is applied to describe the flow field scattered from
the particles, and the Cartesian basis is used in the analysis of the
interaction of the flow with the walls.  The key result of our study
is a set of transformation formulas for conversion between the
spherical and Cartesian representations.  In our algorithm the
expansion of the flow field into the basis fields is combined with the
two-particle superposition approximation for the friction matrix in
order to incorporate slowly convergent lubrication corrections.  Since
the force multipoles induced on particle surfaces are included to
arbitrary order, results of arbitrary accuracy are obtained.

Our paper is organized as follows.  The induced-force formulation of
the problem is described in \S \ref{Multiparticle hydrodynamic
interactions}, and the multipolar representation of the flow in terms
of force multipoles induced on the particles is recalled in \S
\ref{Force-multipole expansion}.  Our main theoretical results are
outlined in \S \ref{Section on Cartesian representation} and \S
\ref{Cartesian representation of wall Green matrix G}.  The Cartesian
basis set of Stokes flows is defined in \S \ref{Section on Cartesian
representation}, along with the transformation relations between the
Cartesian and spherical bases, displacement theorems for the Cartesian
basis fields and expressions for the wall-reflection matrix.  These
essential elements are combined in \S \ref{Cartesian representation of
wall Green matrix G} to evaluate the wall contribution to the Green's
matrix.  The numerical implementation of our method is outlined in \S
\ref{Numerical algorithm}.  Examples of numerical results (for a
single particle, two particles and many-particle systems) are provided
in \S \ref{Results}.  The multiparticle results have been selected to
illustrate the crossover between the quasi-two-dimensional and
three-dimensional behavior of the friction matrix as a function of the
interparticle distance.

Since the full description of the theory underlying our algorithm
requires more space, this paper outlines only the most important
elements and lists the crucial results indispensable for the numerical
implementation.  The details of our theoretical analysis and a more
complete description of the algorithm are presented in a separate
publication \cite[][]{Bhattacharya-Blawzdziewicz-Wajnryb:2005a},
hereafter referred to as Ref.\ I.

\section{Multiparticle hydrodynamic interactions}
\label{Multiparticle hydrodynamic interactions}

\subsection{Hydrodynamic resistance}
\label{section on hydrodynamic resistance}

We consider a suspension of $N$ spherical particles of the radius $a$
in a creeping flow between two parallel planar walls.  The no-slip
boundary conditions are assumed on the particles and on the walls.
The walls are at the positions $z=0$ and $z=H$, where $H$ is the
separation between walls, and $\br=(x,y,z)$ are the Cartesian
coordinates.  The position of the center of particle $i$ is denoted by
$\bR_i=(X_i,Y_i,Z_i)$, the translational and rotational particle
velocities are denoted by $\bU_i$ and $\bOmega_i$, and the external
forces and torques acting on the particle are denoted by $\totForce_i$
and $\totTorque_i$.

We focus on a system of spheres undergoing translational and
rotational rigid-body motion with no external flow. As in an unbounded
space, the particle dynamics in the system is characterized by the
resistance matrix
\begin{equation}
\label{resistance matrix}
\resistanceMatrix_{ij}=
\left[
  \begin{array}{cc}
    \resistanceMatrixTT_{ij}&\resistanceMatrixTR_{ij}\\
    \resistanceMatrixRT_{ij}&\resistanceMatrixRR_{ij}
  \end{array}
\right],
\qquad i,j=1,\ldots,N,
\end{equation}
defined by the linear relation
\begin{equation}
\label{resistance relation}
\left[
  \begin{array}{c}
\totForce_i\\
\totTorque_i
  \end{array}
\right]
=
\sum_{j=1}^N
\left[
  \begin{array}{cc}
    \resistanceMatrixTT_{ij}&\resistanceMatrixTR_{ij}\\
    \resistanceMatrixRT_{ij}&\resistanceMatrixRR_{ij}
  \end{array}
\right]
\bcdot
\left[
  \begin{array}{c}
     \bU_j\\
     \bOmega_j
  \end{array}
\right]
\end{equation}
between the translational and rotational particle velocities and the
forces and torques.  The dot in equation \refeq{resistance relation}
denotes the matrix multiplication and contraction of the Cartesian
tensorial components of the resistance matrix.  

\subsection{Induced-force formulation}
\label{Induced-force formulation}

The effect of the suspended particles on the surrounding fluid can be
described in terms of the induced force distributions on the particle
surfaces
\begin{equation}
\label{induced forces}
\bF_i(\br)=a^{-2}\delta(r_i-a)\bff_i(\br),
\end{equation}
where 
\begin{equation}
\label{definition of r_i}
\br_i=\br-\bR_i
\end{equation}
and $r_i=|\br_i|$.  By definition of the induced force, the flow field
\begin{equation}
\label{flow field produced by induced forces}
\bv(\br)=\sum_{i=1}^N
  \int\bT(\br,\br')\bcdot\bF_i(\br')\diff\br'
\end{equation}
is identical to the velocity field in the presence of the moving
particles
\cite[][]{Cox-Brenner:1967,Mazur-Bedeaux:1974,Felderhof:1976b}.  Here
\begin{equation}
\label{Green's function}
\bT(\br,\br')=\bT_0(\br-\br')+\bT'(\br,\br')
\end{equation}
is the Green's function for the Stokes flow in the presence of the
boundaries; the Green's function $\bT(\br,\br')$ is decomposed into
the Oseen tensor $\bT_0(\br-\br')$ and the part $\bT'(\br,\br')$ that
describes the flow reflected from the walls.  In equation \refeq{flow
field produced by induced forces} it is assumed that the particles
move with given velocities, but no external flow is imposed.

The resistance relation \refeq{resistance relation} is linked to the
induced-force distributions \refeq{induced forces} through the
expressions
\begin{equation}
\label{force and torque}
   \totForce_i=\int\bF_i(\br)\diff\br,
\qquad
   \totTorque_i=\int\br_i\btimes\bF_i(\br)\diff\br
\end{equation}
for the total force and torque, respectively.  To determine the
resistance matrix \refeq{resistance matrix} we thus need to evaluate
the induced forces \refeq{induced forces} for given translational and
angular velocities of the particles.

\subsection{Boundary-integral equations for the induced forces}
\label{section on boundary-integral equations for the induced forces}

For a system of particles moving with the translational and angular
velocities $\bU_i$ and $\bOmega_i$, the induced-force distribution
\refeq{induced forces} can be obtained from the boundary-integral
equation of the form
\begin{equation}
\label{boundary-integral equation for induced-force density}
[\bZ_i^{-1}\bF_i](\br)
   +\sum_{j=1}^N\int
      [(1-\delta_{ij})\bT_0(\br-\br')
+\bT'(\br,\br')]
      \bcdot\bF_j(\br')\diff\br'
   =\bv_i^{\rm rb}(\br),\qquad\br\in S_i,
\end{equation}
where 
\begin{equation}
\label{rigid-body velocity of drop i}
\bv_i^{\rm rb}(\br)=\bU_i+\bOmega_i\times\br_i
\end{equation}
is the rigid-body velocity field associated with the particle motion,
and $S_i$ is the surface of particle $i$.  In the boundary-integral
equation \refeq{boundary-integral equation for induced-force density},
$\bZ_i$ denotes the one-particle scattering operator that describes
the response of an individual particle to an external flow in an
unbounded space.  This operator is defined by the linear relation
\begin{equation}
\label{definition of operator Z}
\bF_i=-\bZ_i(\incidentVelocity{i}-\bv_i^{\rm rb}),
\end{equation}
where $\incidentVelocity{i}$ is the velocity incident to particle $i$.
For specific particle models, explicit expressions for the operator
$\bZ_i$ are known
\cite[][]{%
Jones-Schmitz:1988,%
Cichocki-Felderhof-Schmitz:1988,%
Blawzdziewicz-Wajnryb-Loewenberg:1999%
}.  

\section{Force-multipole expansion}
\label{Force-multipole expansion}

\subsection{Spherical basis fields}
\label{section on spherical basis fields}

As in a standard force-multipole approach
\cite[][]{%
Cichocki-Felderhof-Hinsen-Wajnryb-Blawzdziewicz:1994,%
Cichocki-Jones-Kutteh-Wajnryb:2000%
}
the boundary-integral equation \refeq{boundary-integral equation for
induced-force density} is transformed into a linear matrix equation by
projecting it onto a spherical basis of Stokes flow.  To this end we
use the reciprocal basis sets defined by
\cite{Cichocki-Felderhof-Schmitz:1988}; we introduce, however, a
slightly different normalization to exploit the full symmetry of the
problem.

The singular and nonsingular spherical basis solutions of Stokes
equations $\sphericalBasisM{lm\sigma}(\br)$ and
$\sphericalBasisP{lm\sigma}(\br)$ (with $l=1,2,\ldots$;
$m=-l,\ldots,l$; and $\sigma=0,1,2$) have the following separable form
in the spherical coordinates $\br=(r,\theta,\phi)$:
\begin{subequations}
\label{spherical basis v +-}
\begin{equation}
\label{spherical basis v -}
\sphericalBasisM{lm\sigma}(\br)
       =\sphericalBasisCoefM{lm\sigma}(\theta,\phi)r^{-(l+\sigma)},
\end{equation}
\begin{equation}
\label{spherical basis v +}
\sphericalBasisP{lm\sigma}(\br)
       =\sphericalBasisCoefP{lm\sigma}(\theta,\phi)r^{l+\sigma-1},
\end{equation}
\end{subequations}
where the coefficients $\sphericalBasisCoefM{lm\sigma}(\theta,\phi)$
and $\sphericalBasisCoefP{lm\sigma}(\theta,\phi)$ are combinations of
vector spherical harmonics with angular order $l$ and azimuthal order
$m$.  This property and the $r$-dependence in equations
\refeq{spherical basis v +-} define the Stokes-flow fields
$\sphericalBasisPM{lm\sigma}(\br)$ up to a normalization constant.
Explicit expressions for the functions
$\sphericalBasisCoefPM{lm\sigma}$ in our present normalization are
given in Appendix \ref{Spherical basis}.  The justification for this choice
of the normalization is discussed in Ref.\ I.

Following \cite{Cichocki-Felderhof-Schmitz:1988} we also introduce the
reciprocal basis fields $\reciprocalSphericalBasisPM{lm\sigma}(\br)$,
defined here by the orthogonality relations of the form
\begin{equation}
\label{orthogonality relations for spherical reciprocal basis}
\langle\deltab{a}\reciprocalSphericalBasisPM{lm\sigma}\mid
    \sphericalBasisPM{l'm'\sigma'}\rangle
    =\delta_{ll'}\delta_{mm'}\delta_{\sigma\sigma'},
\end{equation}
where
\begin{equation}
\label{delta b}
\deltab{a}(\br)=a^{-2}\delta(r-a),
\end{equation}
and 
\begin{equation}
\label{scalar product}
\langle\bA\mid\bB\rangle=\int \bA^*(\br)\boldsymbol{\cdot}\bB(\br)\diff\br.
\end{equation}
The asterisk in the above relation denotes the complex conjugate.  
We note that due to the proper choice of defining properties of the
spherical basis sets, the basis fields $\sphericalBasisM{lm\sigma}$
and $\reciprocalSphericalBasisP{lm\sigma}$ satisfy relation
\cite[][]{Cichocki-Felderhof-Schmitz:1988}
\begin{equation}
\label{flow produced by force multipole}
   \sphericalBasisM{lm\sigma}(\br)=
   \eta\int \bT_0(\br-\br')\deltab{a}(\br')
      \reciprocalSphericalBasisP{lm\sigma}(\br')\diff\br',
\end{equation}
where $\eta$ is the viscosity of the fluid.  Relation \refeq{flow
produced by force multipole} assures that the Lorentz reciprocal
symmetry of Stokes flow is reflected in the symmetry of the resulting
matrix representation of the problem
\cite[][]{Cichocki-Jones-Kutteh-Wajnryb:2000}.

\subsection{Matrix representation}
\label{Section on matrix representation}

The matrix representation of the boundary-integral equation
\refeq{boundary-integral equation for induced-force density} is
obtained by expanding the force distributions induced on each particle
into the induced-force multipoles.  The force multipolar moments of
the force distribution \refeq{induced forces} are defined by the
relation
\begin{equation}
\label{induced force in terms of multipoles}
\bF_i(\br)
   =\sum_{lm\sigma}
      f_i(lm\sigma)
            a^{-2}\delta(r_i-a)
         \reciprocalSphericalBasisP{lm\sigma}(\br_i),
\end{equation}
where $\br_i$ is the relative position \refeq{definition of r_i} with
respect to the particle center.  According to equations \refeq{flow
produced by force multipole} and \refeq{induced force in terms of
multipoles}, the multipolar moments $f_i(lm\sigma)$ are identical
(apart from the trivial factor $\eta$) to the expansion coefficients
of the flow field scattered by an isolated particle in unbounded space
into the singular basis fields $\sphericalBasisM{lm\sigma}$.

To obtain a linear matrix equation for the set of force multipolar
moments $f_i(lm\sigma)$, the multipolar representation of the induced
force \refeq{induced force in terms of multipoles} is inserted into
the boundary-integral equation \refeq{boundary-integral equation for
induced-force density}, and the resulting expression is expanded into
the nonsingular basis solutions \refeq{spherical basis v +}.  In
particular, for the rigid-body velocity field we have
\begin{equation}
\label{expansion of external flow}
\bv_i^{\rm rb}(\br)
   =\sum_{lm\sigma}c_i(lm\sigma)\sphericalBasisP{lm\sigma}(\br_i),
\end{equation}
where the expansion coefficients $c_i(lm\sigma)$ are nonzero only for
$l=1$ and $\sigma=0,1$.  

As the result of this procedure we get the linear force-multipole
equation, which can be written in the form
\begin{equation}
\label{induced force equations in matrix notation}
   \sum_{j=1}^N\sum_{l'm'}
      \GrandMobility_{ij}(lm\mid l'm')
   \bcdot
      \inducedForceMultipole_j(l'm')
      =\externalVelocityCoefficient_i(lm).
\end{equation}
We use here a matrix notation in the three-dimensional linear space
with the components corresponding to the indices $\sigma=0,1,2$ that
identify the tensorial character of the basis flow fields
\refeq{spherical basis v +-}.  Accordingly, the arrays
$\inducedForceMultipole_j(l'm')$ and
$\externalVelocityCoefficient_i(lm)$ have the components
$f_j(l'm'\sigma')$ and $c_i(lm\sigma)$ and the matrix
$\GrandMobility_{ij}(lm\mid l'm')$ has the elements
$\GrandMobilityElement_{ij}(lm\sigma\mid l'm'\sigma')$, where
$\sigma,\sigma'=0,1,2$.  The many-particle resistance matrix
\refeq{resistance matrix} can be obtained by solving equation
\refeq{induced force equations in matrix notation} and projecting the
induced force multipoles onto the total force and torque \refeq{force
and torque}.  Explicit expressions for the resistance matrix in terms
of the generalized friction matrix
$\GrandFriction=\GrandMobility^{-1}$ are given in Appendix
\ref{Transformation vectors X}.

For a wall bounded system the matrix $\GrandMobility$ can be
decomposed into three contributions
\begin{equation}
\label{Grand Mobility matrix}
      \GrandMobility_{ij}(lm\mid l'm')
   =
      \delta_{ij}\delta_{ll'}\delta_{mm'}\Zsingle_i^{-1}(l)
   +
      \GreenFree_{ij}(lm\mid l'm')
   +
      \GreenWall_{ij}(lm\mid l'm').
\end{equation}
The first term $\Zsingle_i^{-1}(l)$ corresponds to the one particle
operator $\bZ_i^{-1}$ in equation \refeq{boundary-integral equation for
induced-force density}.  Accordingly, the matrix $\Zsingle_i(l)$
relates the force multipoles $\inducedForceMultipole_i(l'm')$ induced
on particle $i$ to the coefficients in the expansion of the flow field
incoming to this particle into the nonsingular spherical basis fields
\refeq{spherical basis v +}.  By spherical symmetry, this term is
diagonal in the multipolar orders $l$ and $m$, and for rigid spheres
it is explicitly known \cite[][]{Cichocki-Felderhof-Schmitz:1988}.

The Green matrices $\GreenFree_{ij}(lm\mid l'm')$ and
$\GreenWall_{ij}(lm\mid l'm')$ correspond to the integral operators
with the kernels $\bT_0(\br-\br')$ and $\bT'(\br,\br')$ in equation
\refeq{boundary-integral equation for induced-force density}.  As
discussed by \cite{Cichocki-Jones-Kutteh-Wajnryb:2000} and by
\cite{Bhattacharya-Blawzdziewicz-Wajnryb:2005a} the matrix
$\GreenFree_{ij}(lm\mid l'm')$ coincides (apart from the normalization
factors) with the displacement matrix for spherical basis fields,
which is explicitly known \cite[][]{Felderhof-Jones:1989}.  The only
unknown component in expression \refeq{Grand Mobility matrix} is thus
the wall contribution
\begin{equation}
\label{wall Green matrix in terms of v and w}
\GreenWallElement_{ij}(lm\sigma\mid l'm'\sigma')
 =\langle
   \deltab{a}(\br_i)\reciprocalSphericalBasisP{lm\sigma}(\br_i)
\mid
   \wallSphericalBasisM{l'm'\sigma'}(\br_j)
   \rangle,
\end{equation}
where
\begin{equation}
\label{wall flow produced by force multipole}
\wallSphericalBasisM{l'm'\sigma'}(\br)=
   \int \bT'(\br,\br')\deltab{a}(\br')
      \reciprocalSphericalBasisP{l'm'\sigma'}(\br')\diff\br'.
\end{equation}
In the following sections we express this contribution in terms of
the Cartesian basis set of Stokes flows.

\section{Cartesian representation}
\label{Section on Cartesian representation}

The difficulties associated with the evaluation of the matrix
$\GreenWall_{ij}$ in systems with two planar walls stem from the
incompatibility of the spherical basis fields
$\sphericalBasisPM{lm\sigma}$ with the wall geometry.  In particular,
the image representation of a force multipole
\cite[][]{Cichocki-Jones:1998,Bhattacharya-Blawzdziewicz:2002} is
insufficient for a two-wall system, due to the slow convergence of
the multiple-image series.  We propose here an alternative technique,
which relies on a transformation between the spherical basis fields
\refeq{spherical basis v +-} and a Cartesian basis set of Stokes
flows. In the Cartesian representation the flow reflected from the
walls can be obtained in a closed form; thus the difficulties associated
with the multiple-image series are avoided.

According to our Cartesian representation method, the wall
contribution \refeq{wall Green matrix in terms of v and w} to the
matrix $\GrandMobility$ is evaluated by \subfig{i} expanding the
spherical basis flow field $\sphericalBasisM{l'm'\sigma'}(\br_j)$
produced by a force multipole at the position $\br_j$ into the
Cartesian basis; \subfig{ii} solving the two-wall problem in the
Cartesian representation; and \subfig{iii} transforming the resulting
reflected flow back to the spherical basis \refeq{spherical basis
v +} centered at the position $\br_i$.  As a result of this
procedure, the matrix elements \refeq{wall Green matrix in terms of v
and w} are expressed in terms of two-dimensional Fourier integrals
with respect to the lateral coordinates $x,y$.  Our method is outlined
in the following sections.

\subsection{Cartesian basis}
\label{subsection on Cartesian basis}

To describe the flow field between two walls parallel to the $x$--$y$
plane, it is convenient to use a basis set of Stokes flows of the
separable form
\begin{equation}
\label{general form of Cartesian basis}
\CartesianBasisPM{\bk\sigma}(\br)
   =\CartesianBasisCoefPM{\bk\sigma}(z)\e^{\im\bk\bcdot\brho\pm kz}
\end{equation}
that is consistent with the wall geometry.  Here
\begin{equation}
\label{rho}
\brho=x\ex+y\ey
\end{equation}
is the projection of the vector $\br$ onto the $x$--$y$ plane, 
\begin{equation}
\label{wave vector}
\bk=k_x\ex+k_y\ey
\end{equation}
is the corresponding two-dimensional wave vector, and $k=|\bk|$.  By
analogy to the spherical basis \refeq{spherical basis v +-}, there
exist three types of solutions $\sigma=0,1,2$ for each $\bk$.  These
solutions involve a potential flow, a flow with nonzero vorticity, and
a pressure-driven flow.  Explicit expressions for the coefficients
$\CartesianBasisCoefPM{\bk\sigma}(z)$ are given in Appendix
\ref{Appendix on Cartesian basis fields}.  To achieve a substantial
simplification of our final results, the relative amplitudes of the
three components in the basis fields \refeq{general form of Cartesian
basis} have been carefully chosen, as discussed in Ref.\ I.

\subsection{Transformation relations}
\label{Transformation relations}

The transformation relations between the spherical and Cartesian sets
of solutions of Stokes equations can be expressed by the formulas
\begin{eqnarray}
\label{spherical - in Cartesian -- same point}
\sphericalBasisM{lm\sigma}(\br)=\int\diff\bk'\sum_{\sigma'}
   \CartesianBasisPM{\bk'\sigma'}(\br)
      \TransformationElementCS{\pm-}(\bk',lm;\sigma'\mid\sigma),\nonumber\\
\qquad \pm z<0,
\end{eqnarray}
\begin{equation}
\label{Cartesian - in spherical -- same point}
\CartesianBasisPM{\bk\sigma}(\br)=\sum_{l'm'\sigma'}
   \sphericalBasisP{l'm'\sigma'}(\br)
      \TransformationElementSC{+\pm}(l'm',\bk;\sigma'\mid\sigma).
\end{equation}
As demonstrated in Ref.\ I, the transformation matrices
$\TransformationCS{\pm-}(\bk,lm)$ and
$\TransformationSC{+\pm}(lm,\bk)$ have the factorized form
\begin{subequations}
\label{factorization of transformations SC and CS}
\begin{equation}
\label{factorization of transformation SC}
\TransformationSC{+\pm}(lm,\bk)
   =(-\im)^m(2\pi k)^{-1/2}\e^{-\im m\psi}
      \Kmatrix(k,l)\bcdot\tildeTransformationSC{+\pm}(lm),
\end{equation}
\begin{equation}
\label{factorization of transformation CS}
\TransformationCS{\pm-}(\bk,lm)
   =\im^m(2\pi k)^{-1/2}\e^{\im m\psi}
      \tildeTransformationCS{\pm-}(lm)\bcdot\Kmatrix(k,l),
\end{equation}
\end{subequations}
where $\psi$ is the polar angle in the Fourier space, 
\begin{equation}
\label{matrix K}
\KmatrixElement(k,l;\sigma\mid\sigma')=\delta_{\sigma\sigma'}k^{l+\sigma-1},
\end{equation}
and the matrices $\tildeTransformationSC{+\pm}(lm)$ and
$\tildeTransformationCS{\pm-}(lm)$ are independent of the wave vector
$\bk$.  Due to the proper choice of the spherical and Cartesian
fields, the transformation matrices
$\tildeTransformationSC{+\pm}(lm)$ and
$\tildeTransformationCS{\pm-}(lm)$ have a simple symmetric form
\begin{subequations}
\label{form of transformation SC and CS}
\begin{equation}
\label{form of transformation SC}
   \tildeTransformationSC{++}=[\tildeTransformationSC{--}]^\dagger=
\left[
\begin{array}{ccc}
      a&b&c\\
      0&2a&2b\\
      0&0&4a
\end{array}
\right],
\end{equation}
\begin{equation}
   \tildeTransformationSC{+-}=[\tildeTransformationCS{+-}]^\dagger=
(-1)^{l+m}\left[
\begin{array}{ccc}
      c&b&a\\
      -2b&-2a&0\\
      4a&0&0
\end{array}
\right],
\end{equation}
\end{subequations}
where the dagger denotes the Hermitian conjugate.
The three independent scalar coefficients in equations \refeq{form of
transformation SC and CS} are
\begin{subequations}
\label{coefficients a,b,c}
\begin{eqnarray}
\label{coefficient a}
a&=&[4(l-m)!(l+m)!(2l+1)]^{-1/2},\\\nonumber\\
\label{coefficient b}
b&=&2am/l,\\\nonumber\\
\label{coefficient c}
c&=&a\frac{l(2l^2-2l-1)-2m^2(l-2)}{l(2l-1)}.
\end{eqnarray}
\end{subequations}

\subsection{Cartesian displacement formulas}
\label{Cartesian displacement formulas}

In an analysis of the flow between the walls it is convenient to use
the Cartesian basis fields centered at different positions (e.g., the
particle or wall position).  As shown in Ref.\ I, the fields
\refeq{general form of Cartesian basis} centered at different points
$\bR_1$ and $\bR_2$ are related by the displacement formula
\begin{equation}
\label{Cartesian displacement theorems -+}
\CartesianBasisPM{\bk\sigma}(\br_2)=\sum_{\sigma'}
   \CartesianBasisPM{\bk\sigma'}(\br_1)
   \CartesianDisplacementElements{\pm\pm}(\bR_{12},\bk;\sigma'\mid\sigma),
\end{equation}
where $\br_1=\br-\bR_1$, $\br_2=\br-\bR_2$, and
$\bR_{12}=\bR_1-\bR_2$.  Since the shift of the origin of the
coordinate system preserves the behavior of the flow fields
\refeq{general form of Cartesian basis} at infinity, the superscripts
in equation \refeq{Cartesian displacement theorems -+} are either all
positive or all negative.  The displacement matrices
$\CartesianDisplacement{\pm\pm}(\bR_{12},\bk)$ can be factorized as
follows,
\begin{equation}
\label{factorization of Cartesian displacement matrix}
\CartesianDisplacement{\pm\pm}(\bR_{12},\bk)
   =\tildeCartesianDisplacement{\pm\pm}(kZ_{12})\e^{\im\bk\bcdot\brho_{12}},
\end{equation}
where
\begin{equation}
\label{expression for tilde S}
\tildeCartesianDisplacement{--}(kZ)=
\left[
   \begin{array}{ccc}
      1&0&0\\
      0&1&0\\
      -2kZ&0&1
   \end{array}
\right]
\be^{-kZ},\qquad
\tildeCartesianDisplacement{++}(kZ)=
\left[
   \begin{array}{ccc}
      1&0&2kZ\\
      0&1&0\\
      0&0&1
   \end{array}
\right]\be^{kZ},
\end{equation}
and
\begin{equation}
\label{definition of rho_12}
\bR_{12}=\brho_{12}+Z_{12}\ez.
\end{equation}

\subsection{Single-wall reflection matrix}
\label{Single-wall reflection matrix}

The Cartesian basis fields \refeq{general form of Cartesian basis} are
well suited for a description of the interaction of the flow with planar
walls because, due to the translational invariance of the problem, the
lateral Fourier modes with different wave vectors $\bk$ do not couple.
The effect of a single wall on the flow field in the system can be
characterized in terms of the one-wall reflection matrix
$\ZsingleWall$.  To define this quantity we consider Stokes flow in a
fluid bounded by a single wall in the plane
\begin{equation}
\label{position of wall}
z=\Zwall{}.
\end{equation}
The fluid occupies either the halfspace $z>\Zwall{}$ (denoted by
$\halfspace{+}$) or $z<\Zwall{}$ (denoted by $\halfspace{-}$).

The velocity field in the halfspace $\halfspace{\pm}$ can be uniquely
decomposed into the incoming and reflected flows
\begin{equation}
\label{decomposition of flow into incoming and reflected components}
\bv(\br)=\vWin(\br)+\vWout(\br).
\end{equation}
The flow $\vWin(\br)$ is nonsingular in the halfspace
$\halfspace{\mp}$, and the flow $\vWout(\br)$ is nonsingular in the
halfspace $\halfspace{\pm}$.  Thus these flows have the following
expansions in the Cartesian basis,
\begin{subequations}
\label{incoming and reflected flow for one wall}
\begin{equation}
\label{incoming flow}
\vWin(\br)
  =\int\diff\bk\sum_\sigma\cWinCoef{\bk\sigma}
                     \CartesianBasisPM{\bk\sigma}(\rwall{}),
\end{equation}
\begin{equation}
\label{outcoming flow}
\vWout(\br)
  =\int\diff\bk\sum_\sigma \cWoutCoef{\bk\sigma}
                     \CartesianBasisMP{\bk\sigma}(\rwall{}).
\end{equation}
\end{subequations}
Here 
\begin{equation}
\label{r wall}
\rwall{}=\br-\Rwall{}
\end{equation}
denotes the position of the point $\br$ relative to the wall, where
$\Rwall{}=(\Xwall{},\Ywall{},\Zwall{})$ has arbitrary lateral
coordinates $\Xwall{}$ and $\Ywall{}$.  

The single-wall scattering matrix $\ZsingleWall$ relates the expansion
coefficients of the incoming and reflected flows:
\begin{equation}
\label{definition of Z wall}
\cWout(\bk)=-\ZsingleWall
   \bcdot\cWin(\bk),
\end{equation}
where $\cWout(\bk)$ and $\cWin(\bk)$ denote the arrays of expansion
coefficients in equations \refeq{incoming and reflected flow for one
wall}.  For a rigid wall with no-slip boundary
conditions we have
\begin{equation}
\label{expression for Z wall}
\ZsingleWall
   =\left[
       \begin{array}{ccc}
          1&0&0\\
          0&1&0\\
          0&0&1
       \end{array}
    \right],
\end{equation}
as shown in Ref.\ I.  For planar interfaces with other boundary
conditions \cite*[e.g., a surfactant-covered fluid-fluid interface
discussed by][]{Blawzdziewicz-Cristini-Loewenberg:1999} the scattering
matrix can also be obtained.

\section{Evaluation of the wall contribution to  
Green's matrix}
\label{Cartesian representation of wall Green matrix G}

\subsection{Single-wall system}
\label{Single-wall problem}

The transformation, displacement, and reflection matrices described in
\S \ref{Section on Cartesian representation} can be used to construct
the matrix $\GreenWall_{ij}$ for a suspension bounded by a single
planar wall or by two planar walls.  For a single wall, the matrix
\refeq{wall Green matrix in terms of v and w} can be expressed by the
two-dimensional Fourier integral
\begin{equation}
\label{expression for single wall G' -- Fourier integral}
\GreenWall_{ij}(lm\mid l'm')=
   \int\diff\bk\,
   \tilde\Psi_\single(\bk;Z_{i\wall},Z_{\wall j})\e^{\im\bk\bcdot\brho_{ij}}
\end{equation}
with the integrand of the form
\begin{equation}
\label{expression for single wall G' -- integrand}
\tilde\Psi_\single(\bk;Z_{i\wall},Z_{\wall j})=
-\eta^{-1}
   \TransformationSC{+\mp}(lm,\bk)
\bcdot
   \tildeCartesianDisplacement{\mp\mp}(kZ_{i\wall})
\bcdot
   \ZsingleWall
\bcdot
   \tildeCartesianDisplacement{\pm\pm}(kZ_{\wall j})
\bcdot
   \TransformationCS{\pm-}(\bk,l'm'),
\end{equation}
where $Z_{i\wall}=Z_i-Z_{\wall}$ and $Z_{\wall j}=Z_{\wall}-Z_j$
are the vertical offsets between the points $i$ and $j$ and the wall.

The physical interpretation of relation \refeq{expression for single
wall G' -- integrand} is straightforward.  First, the spherical
components of the flow produced by a multipolar force distribution at
point $j$ are transformed by the matrix $\TransformationCS{\pm-}$ into
the Cartesian basis.  The Cartesian components of the velocity field
are propagated by the matrix
$\CartesianDisplacement{\pm\pm}(\bR_{\wall j})$ to the wall, where
they are scattered, as represented by the matrix $\ZsingleWall$.  The
reflected field is propagated by the matrix
$\CartesianDisplacement{\mp\mp}(\bR_{i\wall})$ to the point $i$, and,
finally, the flow is transformed by the matrix
$\TransformationSC{+\mp}$ back into the spherical basis.

Due to symmetry properties of the component matrices (cf.,
relations \refeq{factorization of transformations SC and CS},
\refeq{form of transformation SC and CS}, \refeq{expression for tilde
S}, and \refeq{expression for Z wall}) the wall contribution to the
Green's matrix \refeq{expression for single wall G' -- Fourier integral}
satisfies the Lorentz symmetry
\begin{equation}
\label{reciprocal relations elements of wall Green operator}
\GreenWall_{ij}(lm\mid l'm')
   =\GreenWallCon_{ji}(l'm'\mid lm).
\end{equation}
We note that for the single-wall problem the Fourier integral
\refeq{expression for single wall G' -- Fourier integral} can be
explicitly performed, which yields the image-singularity result
derived by \cite{Cichocki-Jones:1998}.  As discussed in
\S\ref{Numerical algorithm}, both the Fourier representation
\refeq{expression for single wall G' -- Fourier integral} and the
result of \cite{Cichocki-Jones:1998} are used in our algorithm to
accelerate the convergence of the two-wall integrals by a subtraction
of the single-wall contributions.

\subsection{Two-wall system}
\label{Two-wall problem}

The single-wall result presented above can be generalized to the flow
between two parallel walls.  We assume that the walls are in the
planes
\begin{equation}
\label{positions of two walls}
z=\Zlow{}, \qquad z=\Zup{},
\end{equation}
where
\begin{equation}
\label{relative position of walls}
\Zlow{}<\Zup{}.
\end{equation}
The two-wall Green's matrix \refeq{wall Green matrix in terms of v and
w} can be expressed in the form analogous to equations
\refeq{expression for single wall G' -- Fourier integral} and
\refeq{expression for single wall G' -- integrand}, i.e.,
\begin{equation}
\label{expression for two wall G' -- Fourier integral}
\GreenWall_{ij}(lm\mid l'm')=
   \int\diff\bk\,
   \tilde\Psi(\bk;Z_{i\low},Z_{j\low},Z_{\low\up})
   \e^{\im\bk\bcdot\brho_{ij}},
\end{equation}
\begin{equation}
\label{expression for two wall G' -- integrand}
\tilde\Psi(\bk;Z_{i\low},Z_{j\low},Z_{\low\up})=
-\eta^{-1}
   \TSC(lm,\bk)\bcdot\tildeSpw{i}(\bk)
      \bcdot\tildeZW(\bk)
\bcdot
      \tildeSwp{j}(\bk)\bcdot\TCS(\bk,l'm'),
\end{equation}
where the component matrices are given by
\begin{subequations}
\label{two wall transformation matrices}
\begin{equation}
\label{two wall C-S transformation matrix}
   \TCS(\bk,lm)=
\left[
\begin{array}{c}
      \TransformationCS{+-}(\bk,lm)
   \\\\
      \TransformationCS{--}(\bk,lm)
\end{array}
\right],
\end{equation}
\begin{equation}
\label{two wall S-C transformation matrix}
   \TSC(lm,\bk)=
\left[
\begin{array}{cc}
      \TransformationSC{+-}(lm,\bk) & \TransformationSC{++}(lm,\bk)
\end{array}
\right],
\end{equation}
\end{subequations}
\begin{subequations}
\label{two wall displacement}
\begin{equation}
\label{two wall w-p displacement}
   \tildeSwp{j}(\bk)=
\left[
   \begin{array}{cc}
      \tildeCartesianDisplacement{++}(Z_{\low j}\bk)&0
\\\\
      0&\tildeCartesianDisplacement{--}(Z_{\up j}\bk)
   \end{array}
\right],
\end{equation}
\\
\begin{equation}
\label{two wall p-w displacement}
   \tildeSpw{i}(\bk)=
\left[
   \begin{array}{cc}
      \tildeCartesianDisplacement{--}(Z_{i\low}\bk)&0
\\\\
      0&\tildeCartesianDisplacement{++}(Z_{i\up}\bk)
   \end{array}
\right],
\end{equation}
\end{subequations}
and 
\begin{equation}
\label{two wall Z matrix}
   \tildeZW(\bk)=
\left[
   \begin{array}{cc}
      \ZsingleWall^{-1}&\tildeCartesianDisplacement{++}(Z_{\low\up}\bk)
\\\\
      \tildeCartesianDisplacement{--}(Z_{\up\low}\bk)&\ZsingleWall^{-1}
   \end{array}
\right]^{-1}.
\end{equation}

The physical interpretation of equation \refeq{expression for two wall
G' -- integrand} is similar to the interpretation of relation
\refeq{expression for single wall G' -- integrand}, except that two
separate sets of expansion coefficients are now used for the flow
field incoming to the lower and upper walls, which is reflected in the
corresponding block structure of matrices \refeq{two wall
transformation matrices}--\refeq{two wall Z matrix}.  The matrix
$\TCS(\bk,l'm')$ transforms the field produced by a force multipole at
the position $\bR_j$ into the Cartesian basis; the basis fields
$\CartesianBasisP{\bk\sigma}$ are used in the region $Z<z_j$ and the
basis fields $\CartesianBasisM{\bk\sigma}$ in the region $Z>z_j$,
consistent with relation \refeq{spherical - in Cartesian -- same
point}. The Cartesian fields are then translated to the positions of
the respective walls by the matrix $\tildeSwp{j}(\bk)$.  The matrix
$\tildeZW(\bk)$, defined by equation \refeq{two wall Z matrix},
describes the interaction of the flow field with both walls.  We note
that this matrix involves the displacements matrices
$\tildeCartesianDisplacement{--}(Z_{\up\low}\bk)$ and
$\tildeCartesianDisplacement{++}(Z_{\low\up}\bk)$, which correspond
to the propagation of the flow field between the walls in the
multiple-reflection process.  After reflection from the walls is
completed, the matrix $\tildeSpw{i}(\bk)$ propagates the flow field
to the target point $i$, and the matrix $\TSC(lm,\bk)$ transforms it
back into the spherical representation.

Due to the symmetries of the $3\times3$ transformation and displacement
matrices, the corresponding symmetry relations
\begin{subequations}
\label{symmetries of two wall matrices}
\begin{equation}
\label{symmetry of two wall transformation matrices}
\TCS(\bk,lm)=[\TSC(lm,\bk)]^\dagger,
\end{equation}
\begin{equation}
\label{symmetry of two wall displacement matrices}
\tildeSwp{i}(\bk)=[\tildeSpw{i}(\bk)]^\dagger,
\end{equation}
\begin{equation}
\label{symmetry of two wall scattering matrices}
\tildeZW(\bk)=[\tildeZW(\bk)]^\dagger
\end{equation}
\end{subequations}
are satisfied by the matrices \refeq{two wall transformation
matrices}--\refeq{two wall Z matrix}.  Equation \refeq{expression for
two wall G' -- integrand} thus implies that the Green matrix
\refeq{expression for two wall G' -- Fourier integral} satisfies the
Lorentz symmetry \refeq{reciprocal relations elements of wall Green
operator}.

\section{Numerical implementation}
\label{Numerical algorithm}  

The evaluation of the resistance matrix $\resistanceMatrix^{AB}_{ij}$
from relations given in Appendix \ref{Transformation vectors X}
requires solving the linear algebraic equation \refeq{induced force
equations in matrix notation} for the array of induced-force
multipolar moments in order to obtain the generalized friction
coefficients $\GrandFrictionElement_{ij}(lm\sigma\mid l'm'\sigma')$.
In expression \refeq{Grand Mobility matrix} for the matrix
$\GrandMobility_{ij}$ the single particle scattering matrix
$\Zsingle_i$ and the unbounded-space Green's matrix $\GreenFree_{ij}$
are known explicitly
\cite[][]{Felderhof-Jones:1989,Cichocki-Felderhof-Schmitz:1988}.  The
remaining term---the two-wall contribution $\GreenWall_{ij}$---is
evaluated numerically, using relations \refeq{expression for two wall
G' -- Fourier integral}--\refeq{two wall Z matrix} along with our
expressions for the Cartesian displacement matrices \refeq{expression
for tilde S}, the transformation matrices \refeq{form of
transformation SC and CS}, and the single-wall scattering matrix
\refeq{expression for Z wall}.

Taking into account the structure \refeq{factorization of
transformations SC and CS} of the transformation matrices
$\TransformationSC{+\mp}$ and $\TransformationCS{\pm-}$, the
angular integral in equation \refeq{expression for single wall G' --
Fourier integral} can be performed analytically. The integration
yields the result in the form of a Hankel transform of the order
$m'-m$.  Accordingly, only a one-dimensional integral in equation
\refeq{expression for two wall G' -- Fourier integral} has to be
performed numerically.  The numerical integration is straightforward
when the lateral separation between particles $i$ and $j$ is small
compared to the wall separation.  For large interparticle separations
$\rho_{ij}$, however, the integration is more difficult due to the
oscillatory behavior of the integrand.

To avoid numerical integration of a highly oscillatory function, the
Fourier amplitude in \refeq{expression for two wall G' -- Fourier
integral} is decomposed
\begin{equation}
\label{decomposition of integrand}
\tilde\Psi(\bk)=\tilde\Psi_\low(\bk)+\tilde\Psi_\up(\bk)+\delta\tilde\Psi(\bk)
\end{equation}
into the superposition of the single-wall contributions
$\tilde\Psi_\low$ and $\tilde\Psi_\up$, and the remaining part
$\delta\tilde\Psi$ representing hydrodynamic interactions between the
walls.  From an analysis of expression \refeq{expression for single
wall G' -- integrand} we find that the magnitude of the single-wall
Fourier amplitudes $\tilde\Psi_\low(\bk)$ and $\tilde\Psi_\up(\bk)$ for
large $k$ is
\begin{equation}
\label{asymptotic behavior of integrand for wall alpha}
\tilde\Psi_\alpha(\bk)\sim\e^{-k\imageOffset_{ij}^{(\alpha)}},
\qquad\alpha=\low,\up,
\end{equation}
 where $\imageOffset_{ij}^{(\alpha)}$ is the vertical offset between
 the point $i$ and the reflection of point $j$ in the wall $\alpha$.
 In contrast, the large-$k$ behavior of the wall-interaction part of
 Fourier amplitude \refeq{decomposition of integrand} is
\begin{equation}
\label{asymptotic behavior of wall-interaction integrand}
\delta\tilde\Psi(\bk)\sim\e^{-k\tilde\imageOffset_{ij}},
\end{equation}
where
\begin{equation}
\label{distance to second image}
\tilde\imageOffset_{ij}=2H-|Z_{ij}|>H.
\end{equation}
The lengthscale $\tilde\imageOffset_{ij}$ equals the vertical offset
$|Z_i-Z_j''|$ between the target point $i$ and the closer of the two
second-order images of the source point $j$.  Since
$\delta\tilde\Psi(\bk)$ decays on the wave-vector scale set by the
distance between the walls
$H>\min(\imageOffset_{ij}^{(\low)},\imageOffset_{ij}^{(\up)})$, a
smaller number of oscillations of the Fourier modes contributes to the
integral after the single-wall terms have been subtracted.

In our algorithm, the short-range function $\delta\tilde\Psi(\bk)$ is
integrated numerically.  The one-particle contributions
$\tilde\Psi_\low(\bk)$ and $\tilde\Psi_\up(\bk)$ are evaluated
analytically, using the explicit image-representation expressions
derived in \cite{Cichocki-Jones:1998}.  In this way we avoid
integrating a highly oscillatory function when the particles are close
to a wall.  The procedure can be further improved, either by
subtracting several terms associated with higher-order wall
reflections of the source multipole
\cite[][]{Bhattacharya-Blawzdziewicz:2002}, or by using asymptotic
formulas for the integrals \refeq{expression for two wall G' --
Fourier integral}.  We have recently derived a complete set of such
expressions, which will be presented in a separate publication.

In order to improve convergence with the order $\lmax$ of the
multipoles included in the calculation we employ a standard technique,
originally introduced by \cite{Durlofsky-Brady-Bossis:1987}.
Accordingly, the lubrication forces that cause a slow convergence of
the results with $\lmax$ are included in the friction matrix using a
superposition approximation.  Both, the interparticle and
particle-wall lubrication corrections are included in this way.
Following the implementation of this method by
\cite{Cichocki-Jones-Kutteh-Wajnryb:2000} for a single wall problem,
we represent the elements of resistance matrix
$\resistanceMatrix_{ij}$ in the form
\begin{equation} 
\label{lubrication superposition for resistance matrix}
\resistanceMatrix_{ij}=
    \resistanceMatrix^{{\rm sup},2}_{ij}
      +\resistanceMatrix^{{\rm sup},\wall}_{ij}
      +\Delta\resistanceMatrix_{ij}.
\end{equation}
Here $\resistanceMatrix^{{\rm sup},2}_{ij}$ denotes the superposition
of two-particle resistance matrices evaluated for isolated particle
pairs in the unbounded space, and 
\begin{equation}
\label{particle-wall superposition}
\resistanceMatrix^{{\rm sup},\wall}_{ij}
   =\delta_{ij}\sum_{\alpha=\low,\up}\resistanceMatrix^{\alpha}_{i}(i).
\end{equation}
is the superposition of one-particle contributions in the presence of
individual walls.  The one-particle contributions can be evaluated
using a series expansion of resistance coefficients in inverse powers
of particle-wall separation \cite[][]{Cichocki-Jones:1998} in
combination with the appropriate lubrication results
\cite[][]{Kim-Karrila:1991}.  The two-particle superposition
contributions $\resistanceMatrix^{{\rm sup},\wall}_{ij}$ are evaluated
in a similar way.  The convergence with the multipolar truncation
order $\lmax$ for the quantity $\Delta\resistanceMatrix_{ij}$ is fast;
some convergence tests are presented in Ref.\ I.

In the present implementation of our method, the numerical cost scales
as $O(N^3)$ with the number of particles $N$, because the linear
equation \refeq{Grand Mobility matrix} is solved by inversion of the
matrix $\GrandMobility$.  However, the numerical efficiency of our
algorithm can be substantially improved by applying fast-multipole or
PPPM acceleration methods in combination with asymptotic expressions
for the elements of the matrix $\GrandMobility$.

\begin{figure}
\includegraphics{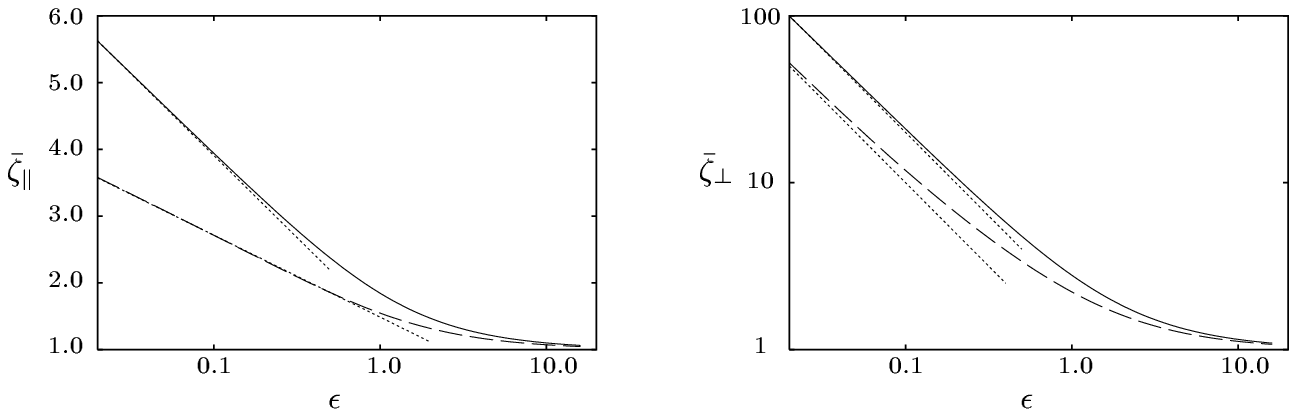}
\input{figtex/singleParticle-cap}
\end{figure}

\section{Results}
\label{Results}

In this section we present a set of numerical results for hydrodynamic
interactions in systems of spherical particles confined between two
parallel planar walls.  Our goal is both, to illustrate typical
behavior of the hydrodynamic friction matrix for particles in the
confined region, and to demonstrate the capabilities of our numerical
algorithm.  The results for a single particle and for pairs of
particles, shown in figures \ref{single-particle friction
coefficients}--\ref{cross terms vertical}, were obtained using the
multipolar approximation with the truncation at the order $\lmax=12$.
This truncation is sufficient to obtain results with the accuracy
better than the resolution of the plots, even for the smallest wall
separation $H$ considered.  The multi-particle results in figures
\ref{rigid-body resistance}--\ref{torque on particles in chain} were
obtained using $\lmax=8$.

\subsection{Single particle}
\label{Single particle}

In figure \ref{single-particle friction coefficients}, the lateral and
vertical friction coefficients
\begin{equation}
\label{single particle lateral and vertical friction matrix}
\onePartResistanceLat
   =
      \resistanceMatrixElement^{\transl\transl\,xx}_{11}
   =
      \resistanceMatrixElement^{\transl\transl\,yy}_{11},
\qquad
\onePartResistanceTrans
   =\resistanceMatrixElement^{\transl\transl\,zz}_{11}
\end{equation}
are shown for a single particle at the center and off-center positions
\refstepcounter{equation}
$$
\label{center and off-center positions}
h=\half H,\qquad h=\third H,
\eqno{(\theequation{\mathit{a},\mathit{b}})}
$$ 
where $h$ is the distance of the particle from the lower wall.  The
results are normalized by the Stokes friction coefficient
$\StokesResistance=6\upi\eta a$,
\begin{equation}
\label{normalized lateral and vertical single-particle friction}
\bar\onePartResistanceLat=\onePartResistanceLat/\StokesResistance,
\qquad
\bar\onePartResistanceTrans=\onePartResistanceTrans/\StokesResistance,
\end{equation}
and are plotted versus the normalized gap 
\begin{equation}
\label{particle-wall gap normalized by a}
\PWgap=h/a-1
\end{equation}
between the particle and the lower wall.  As expected, for small
values of the gap, the lateral and vertical resistance coefficients
approach the asymptotic lubrication behavior (in figure
\ref{single-particle friction coefficients} indicated by dotted
lines).  For $h=\half H$ the lubrication behavior is
\begin{subequations}
\label{one-particle lubrication behavior}
\begin{equation}
\label{one-particle lubrication behavior for middle position}
\bar\onePartResistanceLat
   =-\textstyle\frac{16}{15}\log \PWgap +C(\half),
      \qquad
\bar\onePartResistanceTrans
   =2\PWgap^{-1},
\end{equation}
where the singular terms correspond to the superposition of two
particle--wall lubrication regions \cite[cf.\ lubrication expressions
given by][]{Cichocki-Jones:1998}.  For the off-center position
$h=\third H$ there is only one lubrication region, thus
\begin{equation}
\label{one-particle lubrication behavior for third position}
\bar\onePartResistanceLat
   =-\textstyle\frac{8}{15}\log \PWgap +C(\third),
      \qquad
\bar\onePartResistanceTrans
   =\PWgap^{-1}.
\end{equation}
\end{subequations}
A comparison of the numerical results shown in figure
\ref{single-particle friction coefficients} with the asymptotic
behavior \refeq{one-particle lubrication behavior} yields
$C(\half)=1.45$ and $C(\third)=1.49$.  

We note that our one-particle
results agree with the numerical calculation by
\cite{Ganatos-Weinbaum-Pfeffer:1980,Ganatos-Pfeffer-Weinbaum:1980} and
with our earlier results \cite[][]{Bhattacharya-Blawzdziewicz:2002-C1}
obtained by an image-representation method
\cite[][]{Bhattacharya-Blawzdziewicz:2002}.

\begin{figure}
\includegraphics{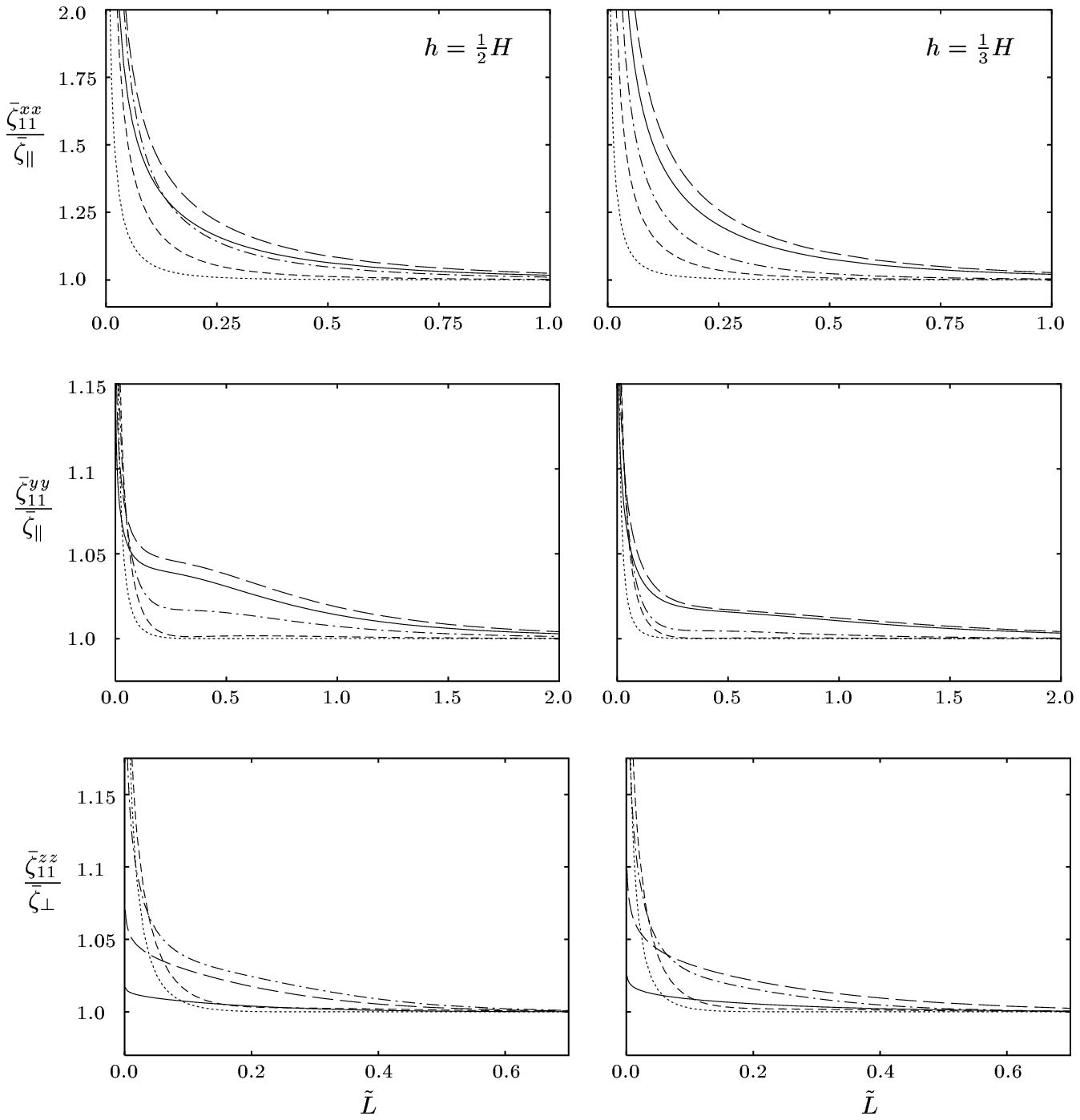}
\input{figtex/selfdia-horizontal-cap}
\end{figure}
\begin{figure}
\includegraphics{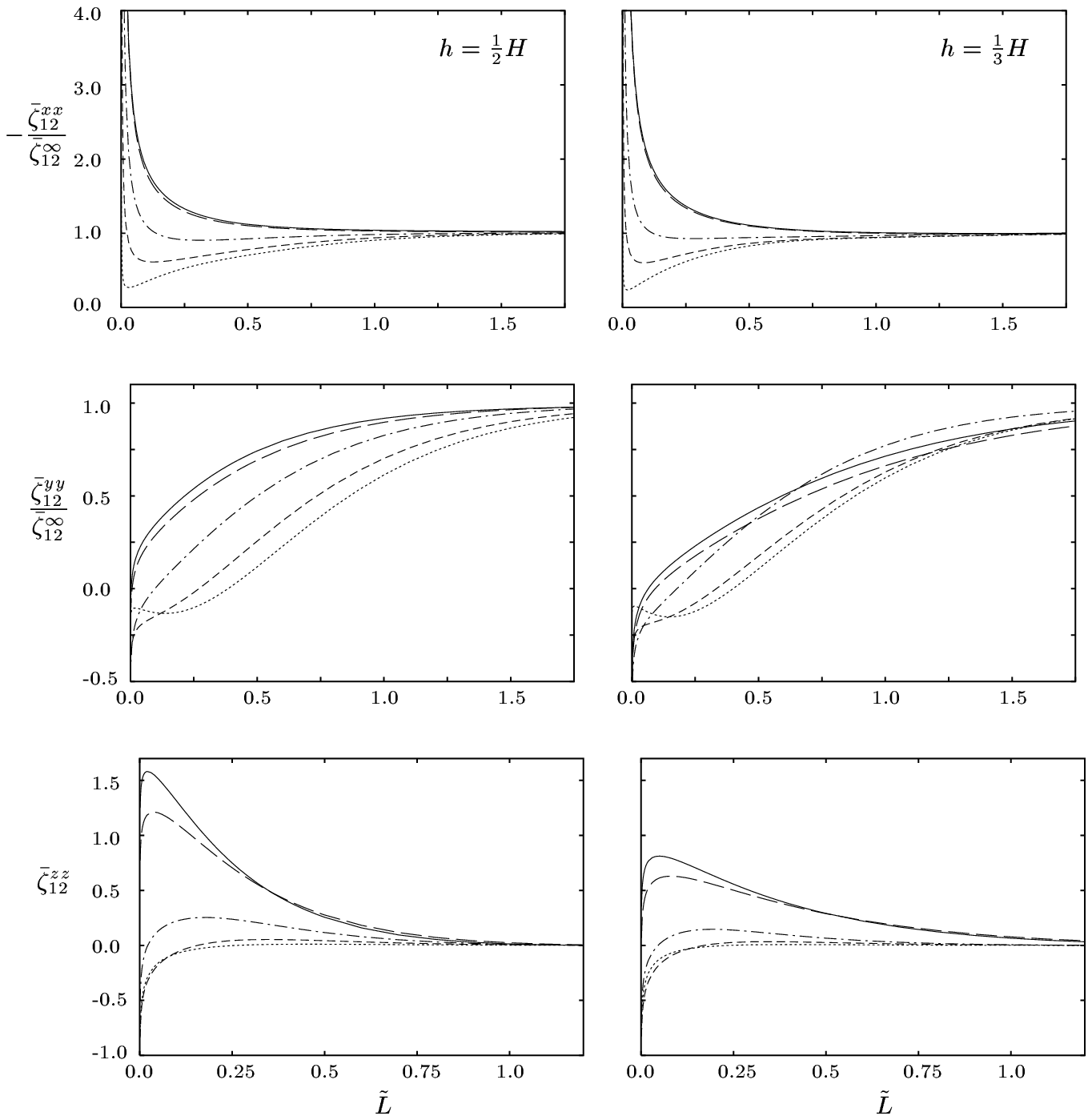}
\input{figtex/mutualdia-horizontal-cap}
\end{figure}
\begin{figure}
\begin{center}
\includegraphics{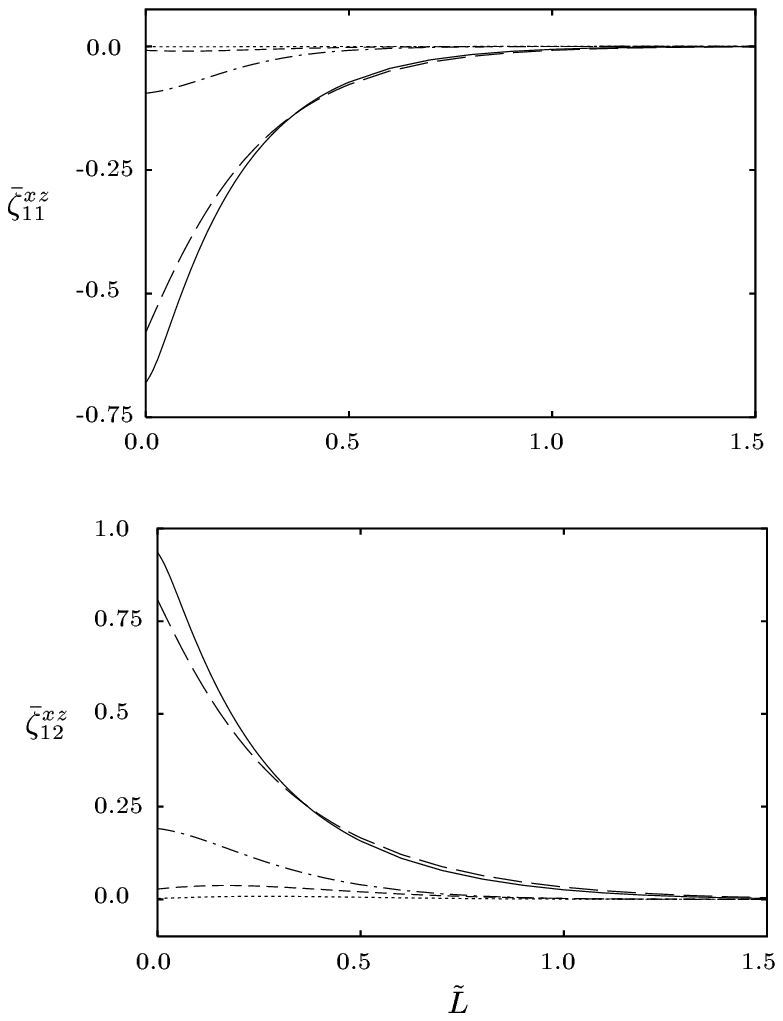}
\end{center}
\input{figtex/bothcross-horizontal-cap}
\end{figure}
\begin{figure}
\begin{center}
\includegraphics{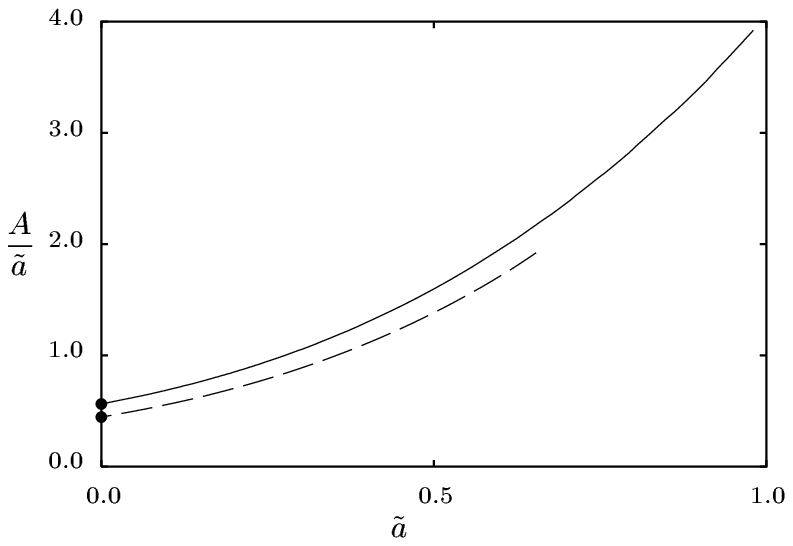}
\end{center}
\input{figtex/horizontalDecayAmplitude-cap}
\end{figure}
\subsection{Two particles}
\label{Two particles}

Sample results for the translational components of the two-particle
resistance matrix
\begin{equation}
\label{normalized two-particle resistance matrix}
\twoParticleResistanceNorm_{ij}^{\alpha\beta}
   =\resistanceMatrixElement_{ij}^{\transl\transl\,\alpha\beta}
    /\StokesResistance,\qquad i,j=1,2,
\end{equation}
(where $\alpha,\beta=x,y,z$) are presented in figures \ref{self
diagonal terms horizontal}--\ref{cross terms vertical}.  The results
in figures \ref{self diagonal terms horizontal}--\ref{horizontal cross
terms} are shown for horizontal particle configurations $h_1=h_2=h$,
where $h_i$ is the distance of particle $i$ from the lower wall.  As
for a single sphere, we consider the center and off-center positions
\refeq{center and off-center positions}.  The relative horizontal
displacement of the particles is
\begin{equation}
\label{relative position along axis x}
\brho_{12}=\rho_{12}\ex.
\end{equation}

To emphasize the crossover between the three-dimensional behavior for
$\rho_{12}\ll H$ and a quasi-two-dimensional behavior for
$\rho_{12}\gg H$, we discuss our results in terms of the dimensionless
variables scaled by the distance between the walls $H$,
\refstepcounter{equation}
$$
\label{variables scaled by H}
\rhoH=\rho_{12}/H,\qquad \aH=a/H,\qquad \PPgap=\rhoH-2\aH.
\eqno{(\theequation{\mathit{a},\mathit{b},\mathit{c}})}
$$ 
The resistance coefficients in figures \ref{self diagonal terms
horizontal}--\ref{horizontal cross terms} are plotted versus the
dimensionless separation between the particle surfaces $\PPgap$.

\subsubsection*{Self-resistance coefficients}

Figures \ref{self diagonal terms horizontal} and \ref{mutual diagonal
terms horizontal} illustrate the behavior of the diagonal components
of the translational self- and mutual resistance matrices
$\twoParticleResistanceNorm_{11}^{\alpha\alpha}$ and
$\twoParticleResistanceNorm_{12}^{\alpha\alpha}$, respectively, and
figure \ref{horizontal cross terms} shows the off-diagonal elements
$\twoParticleResistanceNorm_{11}^{xz}$ and
$\twoParticleResistanceNorm_{12}^{xz}$.  The remaining coefficients of
the two-particle translational resistance matrix either vanish or can
be related to the above coefficients by symmetry.

The results for the resistance coefficients
$\twoParticleResistanceNorm_{11}^{\alpha\alpha}$ presented in figure
\ref{self diagonal terms horizontal} are scaled by the corresponding
single-particle friction coefficients \refeq{normalized lateral and
vertical single-particle friction}.  For small distances between the
particle surfaces $\PPgap\ll\aH$ the longitudinal resistance
coefficient $\twoParticleResistanceNorm_{11}^{xx}$ is dominated by the
$O(\PPgap^{-1})$ interparticle lubrication friction; the lubrication
behavior of the components transverse to the direction of the line
connecting the particle centers is
$\twoParticleResistanceNorm_{11}^{yy},
\twoParticleResistanceNorm_{11}^{zz}\sim\log\PPgap$.

In the intermediate regime $\rhoH\approx1$ the two-particle friction
matrix undergoes a crossover to a quasi-two-dimensional far-field
asymptotic behavior at large interparticle distances.  A signature of
the crossover is the kink seen in the plot of
$\twoParticleResistanceNorm_{11}^{yy}$ for the particles at the
center position $h=\half H$.  For large interparticle separations
$\rhoH\gg1$, the lateral components of the self-friction matrix
approach the one-particle asymptotic value as
\begin{equation}
\label{large rho behavior of self friction matrix}
\twoParticleResistanceNorm_{11}^{xx}
   \approx\twoParticleResistanceNorm_{11}^{yy}
   =\bar\onePartResistanceLat+O(\rhoH^{-4}),\qquad\rhoH\gg1.
\end{equation}
This result stems from the far-field behavior of the disturbance
velocity produced by the particles.  For the lateral motion the
far-field disturbance decays as $O(\rhoH^{-2})$, as shown in Ref.\ I.
Since the contribution of the second particle to the self-components
of the friction matrix
$\twoParticleResistanceNorm_{11}^{\alpha\alpha}$ involves the field
scattered back to the first particle, the asymptotic behavior
\refeq{large rho behavior of self friction matrix} is obtained.  In
contrast, the disturbance field corresponding to the vertical motion
decays exponentially, which yields an exponential approach of the
vertical component of the friction matrix
$\twoParticleResistanceNorm_{11}^{zz}$ to the one-particle value
$\bar\onePartResistanceTrans$.

\subsubsection*{Mutual resistance coefficients}

An analogous reasoning applied to the mutual components of the
friction matrix yields the asymptotic behavior
\begin{equation}
\label{large rho behavior of mutual friction matrix}
\twoParticleResistanceNorm_{12}^{yy}
   \approx-\twoParticleResistanceNorm_{12}^{xx}
   =A\rhoH^{-2}+O(\rhoH^{-4}),\qquad\rhoH\gg1,
\end{equation}
where the amplitude $A>0$ depends on the size of the particles and on
their vertical positions in the gap.  Note that the sign of the
transverse resistance coefficient
$\twoParticleResistanceNorm_{12}^{yy}$ at large interparticle
distances is opposite to the sign of the corresponding coefficient in
the unbounded space.

The results for $\twoParticleResistanceNorm_{12}^{xx}$ and
$\twoParticleResistanceNorm_{12}^{yy}$ shown in figure \ref{mutual
diagonal terms horizontal} are scaled using expression \refeq{large
rho behavior of mutual friction matrix}, with the amplitude $A$
plotted in figure \ref{horizontal decay amplitude} (discussed below).
Since $\twoParticleResistanceNorm_{12}^{zz}$ decays exponentially for
large $\rhoH$, the results for this component are presented unscaled.
Similar to the results in figure \ref{self diagonal terms horizontal}
for the self-resistance matrix, the diagonal components of the mutual
resistance matrix have a lubrication singularity for particles in
contact, and for $\rhoH=O(1)$ they exhibit a crossover to the
asymptotic $O(\rhoH^{-2})$ far-field behavior in the regime
$\rhoH\gg1$.  The near-field and far-field region are most pronounced
for large values of the particle-wall gap $\PWgap$ (i.e., for
$\aH\ll1$) because of the lengthscale separation.

\subsubsection*{Cross-terms}

The cross-elements of the self- and mutual friction matrix
$\twoParticleResistanceNorm_{11}^{xz}$ and
$\twoParticleResistanceNorm_{12}^{xz}$ are shown (unscaled) in figure
\ref{horizontal cross terms}.  Since for the center particle position
(\ref{center and off-center positions}\textit{a}) these components
vanish by symmetry, the results are presented only for the off-center
configuration (\ref{center and off-center positions}\textit{b}).  The
nonzero values of the cross-resistance coefficients
$\twoParticleResistanceNorm_{11}^{xz}$ and
$\twoParticleResistanceNorm_{12}^{xz}$ arise indirectly, due to the
asymmetry of the flow field scattered from the walls.  Therefore, for
$\PPgap=0$ there is no lubrication singularity.  The cross-resistance
coefficients involve vertical particle motion; thus, for large
interparticle separations $\twoParticleResistanceNorm_{11}^{xz}$ and
$\twoParticleResistanceNorm_{12}^{xz}$ decay exponentially.

\subsubsection*{Amplitude of the far-field asymptotic behavior}

The behavior \refeq{large rho behavior of self friction matrix} and
\refeq{large rho behavior of mutual friction matrix} of the
two-particle resistance coefficients for $\rhoH\gg1$ is consistent
with the asymptotic expressions derived by \cite{Liron-Mochon:1976}
for the far-field flow produced by Stokeslets oriented in the
direction parallel and normal to the walls.  Using the Liron--Mochon
expression and applying the Stokes resistance formula to evaluate
forces acting on small particles in the space between the walls yields
the asymptotic behavior \refeq{large rho behavior of mutual friction
matrix}, with the amplitude given by
\begin{equation}
\label{Liron-Mochon formula}
\frac{A}{\aH}=9\hH_1(1-\hH_1)\hH_2(1-\hH_2)+O(\aH),
\end{equation}
where $\hH_i=h_i/H$.  Figure \ref{horizontal decay amplitude} shows
the dependence of the far-field amplitude $A$ on the dimensionless
particle size $\aH$; the limiting result \refeq{Liron-Mochon formula}
is indicated by circles.

We emphasize that the far-field form of the disturbance flow produced
by particles in a domain bounded by parallel walls, and the
corresponding properties of the resistance matrix are important for
understanding the macroscopic dynamics of suspensions in slit-pore
geometries.  A more detailed analysis of this problem will be given
elsewhere.

\begin{figure}
\begin{center}
\includegraphics{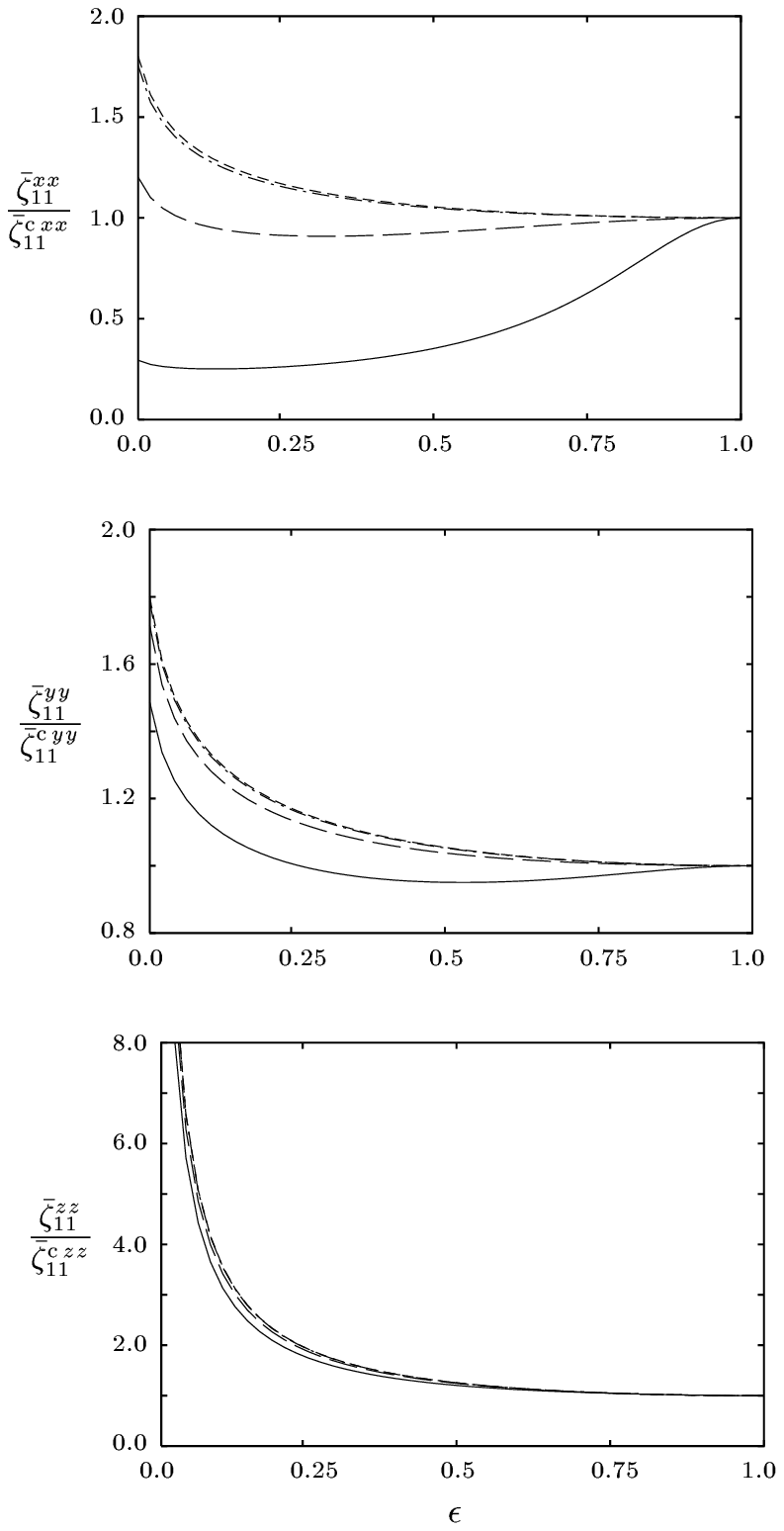}
\end{center}
\input{figtex/selfdia-vertical-cap}
\end{figure}

\begin{figure}
\begin{center}
\includegraphics{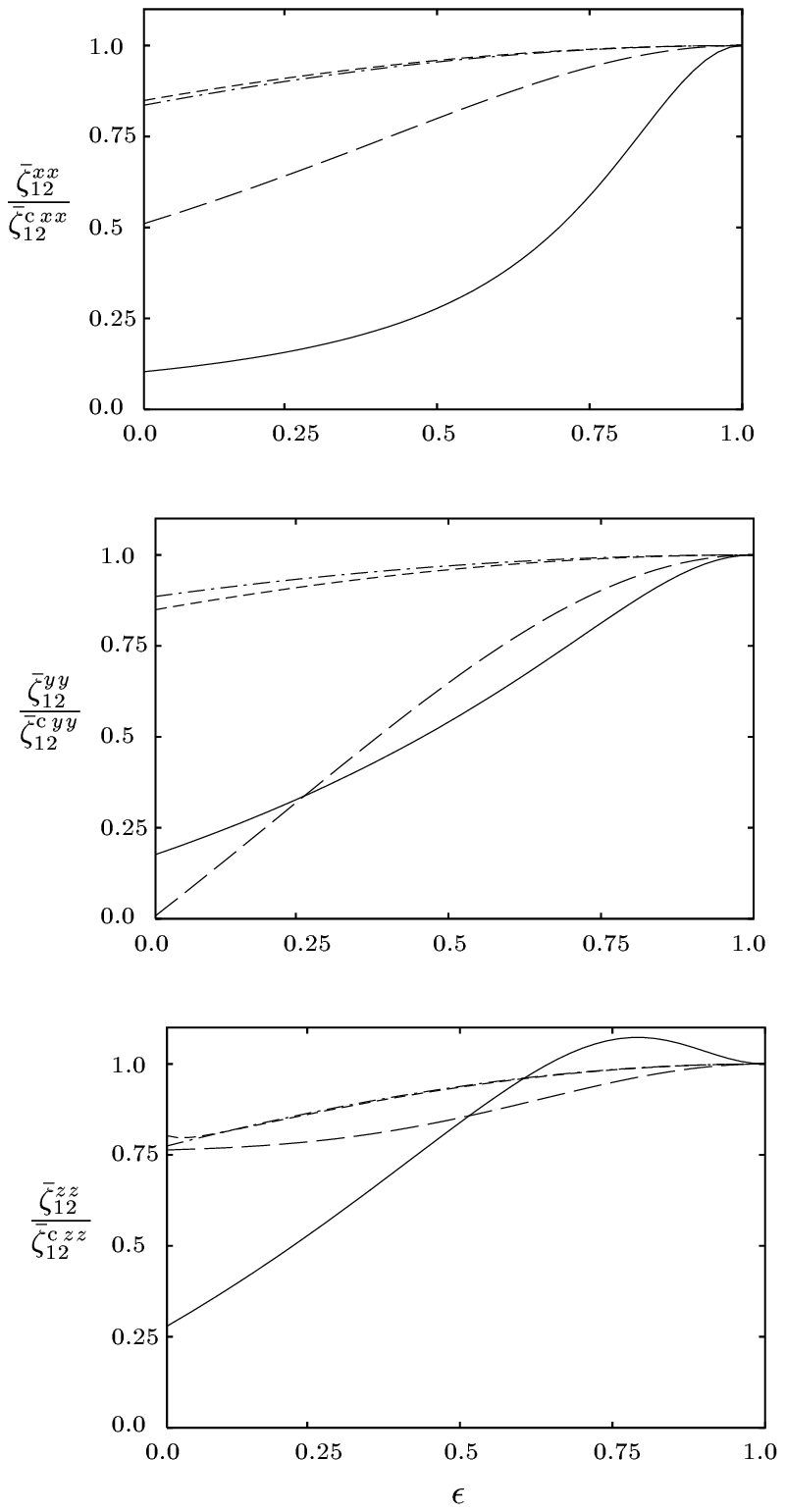}
\end{center}
\input{figtex/mutualdia-vertical-cap}
\end{figure}

\begin{figure}
\begin{center}
\includegraphics{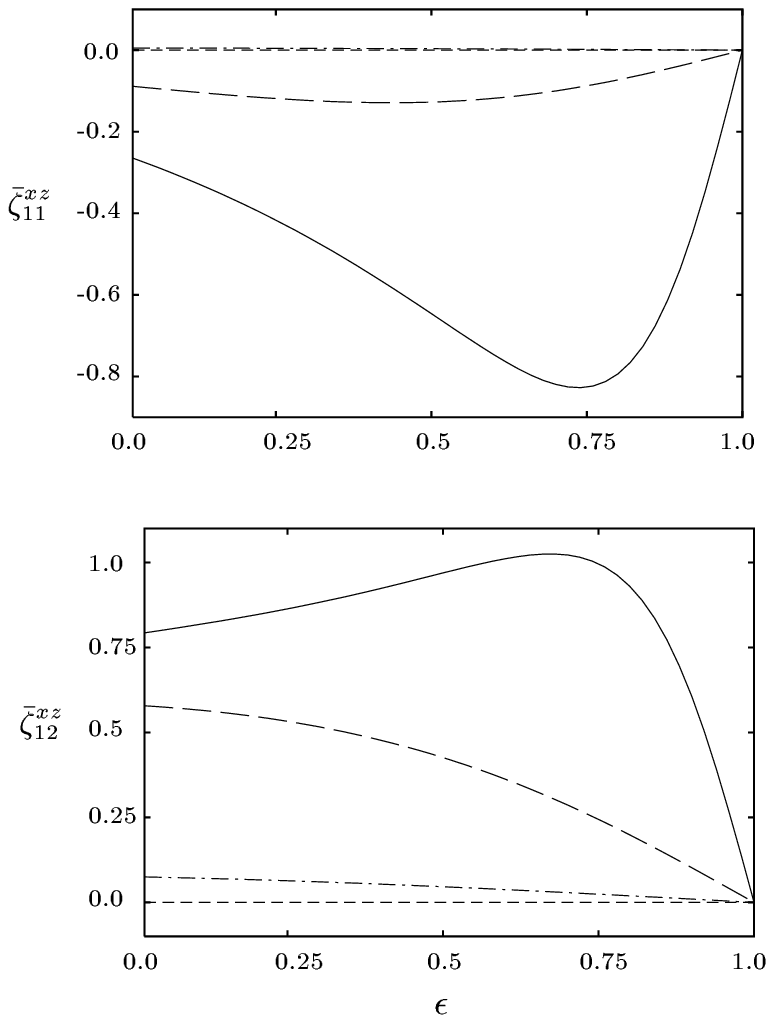}
\end{center}
\input{figtex/bothcross-vertical-cap}
\end{figure}

\begin{table}
\begin{center}
\begin{tabular}{@{}lcccccc@{}}
$\rho/2a$&
$\twoParticleResistanceCenter{11}{xx}$& 
$\twoParticleResistanceCenter{11}{yy}$& 
$\twoParticleResistanceCenter{11}{zz}$& 
$\twoParticleResistanceCenter{12}{xx}$& 
$\twoParticleResistanceCenter{12}{yy}$& 
$\twoParticleResistanceCenter{12}{zz}$\\[3pt]
1.01&15.08&2.31&3.26&-13.75&-0.51&-0.24\\
1.1 &3.35&1.98&2.95&-1.99&-0.117&0.132\\
2.0 &1.93&1.87&2.80&-0.368&0.200&0.117\\
5.0 &1.85&1.85&2.79&-0.064&0.063&0.00015
\end{tabular}
\end{center}
\caption{Normalization factors (\ref{scaling for skew configurations})
for the configurations represented in figures \ref{self diagonal terms
vertical} and \ref{mutual diagonal terms vertical}.}
\label{table of normalization factors}
\end{table}

\subsubsection*{Skew configurations}

So far we have focused on horizontal particle configurations with both
particles at the same distance from the walls.  In figures \ref{self
diagonal terms vertical}--\ref{cross terms vertical} we consider skew
configurations with the vertical positions
\begin{equation}
\label{skew configurations}
h_1=h,\qquad h_2=H-h.
\end{equation}
Figures \ref{self diagonal terms vertical} and \ref{mutual diagonal
terms vertical} show the diagonal components
$\twoParticleResistanceNorm_{11}^{\alpha\alpha}$ and
$\twoParticleResistanceNorm_{12}^{\alpha\alpha}$ of the self- and
mutual resistance matrices, and figure \ref{cross terms vertical}
presents the off-diagonal components
$\twoParticleResistanceNorm_{11}^{xz}$ and
$\twoParticleResistanceNorm_{12}^{xz}$.  The results are plotted
versus the normalized particle-wall gap \refeq{particle-wall gap
normalized by a}.

The diagonal resistance coefficients
$\twoParticleResistanceNorm_{ij}^{\alpha\alpha}$ in figures \ref{self
diagonal terms vertical} and \ref{mutual diagonal terms vertical} are
scaled by the value
\begin{equation}
\label{scaling for skew configurations}
\twoParticleResistanceCenter{ij}{\alpha\alpha} 
   =\twoParticleResistanceNorm_{ij}^{\alpha\alpha}(h=\half H),
\end{equation}
corresponding to the center configuration of the particle pair at a
given lateral separation $\rho$ and wall-to-wall distance $H$.  The
resistance coefficients for the center configuration \refeq{scaling
for skew configurations} have been discussed above; the values of the
normalization factors for the parameter values represented in figures
\ref{self diagonal terms vertical} and \ref{mutual diagonal terms
vertical} are listed in table \ref{table of normalization factors}.

For large lateral interparticle distances, the self-resistance
coefficients, shown in figure \ref{self diagonal terms vertical},
approach the corresponding one-particle results.  We note that due to
the fast asymptotic approach \refeq{large rho behavior of self
friction matrix}, the results for $\rho/2a=5$ essentially coincide
with the one-particle values.  For small particle--wall gaps, the
lateral coefficients $\twoParticleResistanceNorm_{11}^{xx}$ and
$\twoParticleResistanceNorm_{11}^{yy}$ exhibit the logarithmic
lubrication singularity, and the vertical component
$\twoParticleResistanceNorm_{11}^{zz}$ has the $1/\PWgap$ singularity.
The rapid variation of the longitudinal coefficient
$\twoParticleResistanceNorm_{11}^{xx}$ in the regime $\PWgap\approx1$
(center particle positions) for $\rho/2a=1.01$ results from the strong
lubrication interaction between the particles.  The same remark
applies to the mutual longitudinal coefficient
$\twoParticleResistanceNorm_{12}^{xx}$ shown in figure \ref{mutual
diagonal terms vertical}.

According to the results in table \ref{table of normalization
factors}, the mutual resistance coefficients approach zero for large
interparticle distances.  The far-field behavior is consistent with
the asymptotic expression \refeq{large rho behavior of mutual friction
matrix} for the lateral components
$\twoParticleResistanceNorm_{12}^{xx}$ and
$\twoParticleResistanceNorm_{12}^{yy}$ and the asymptotic exponential
decay for the vertical component
$\twoParticleResistanceNorm_{11}^{zz}$.  We note that for small and
moderate interparticle distances there is no simple relation between
the components $\twoParticleResistanceNorm_{12}^{xx}$ and
$\twoParticleResistanceNorm_{12}^{yy}$; however
$\twoParticleResistanceNorm_{12}^{yy}\approx -
\twoParticleResistanceNorm_{12}^{xx}$ for $\rhoH\gg1$, in agreement
with equation \refeq{large rho behavior of mutual friction matrix}.
 
The off-diagonal components $\twoParticleResistanceNorm_{11}^{xz}$ and
$\twoParticleResistanceNorm_{12}^{xz}$, shown unscaled in figure
\ref{cross terms vertical}, are exponentially small for
$\rhoH\gg1$. For both particles at the center of the space between the
walls these components vanish by symmetry.

\begin{figure}
\begin{center}
\includegraphics{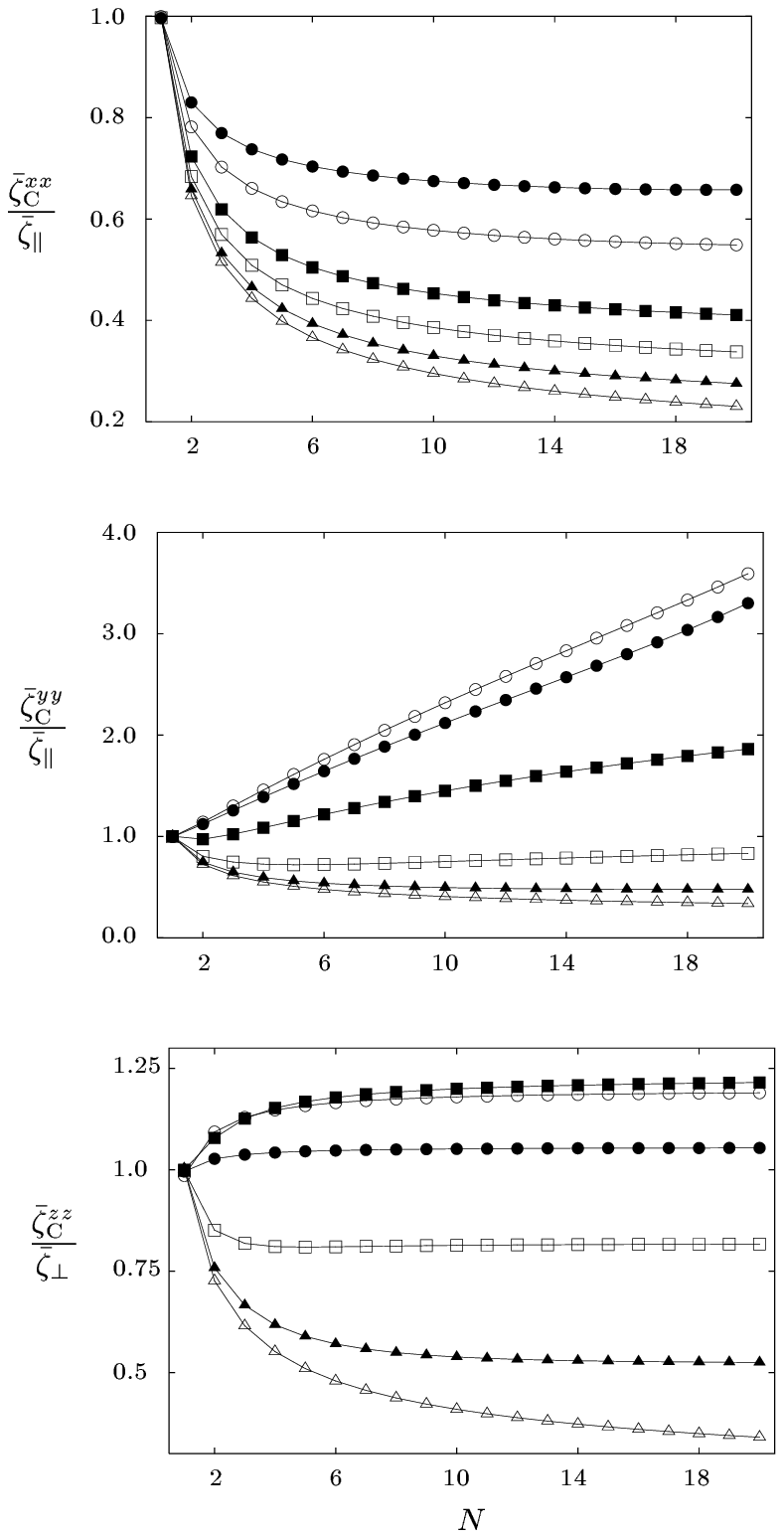}
\end{center}
\input{figtex/polymer-force-cap}
\end{figure}

\begin{figure}
\begin{center}
\includegraphics{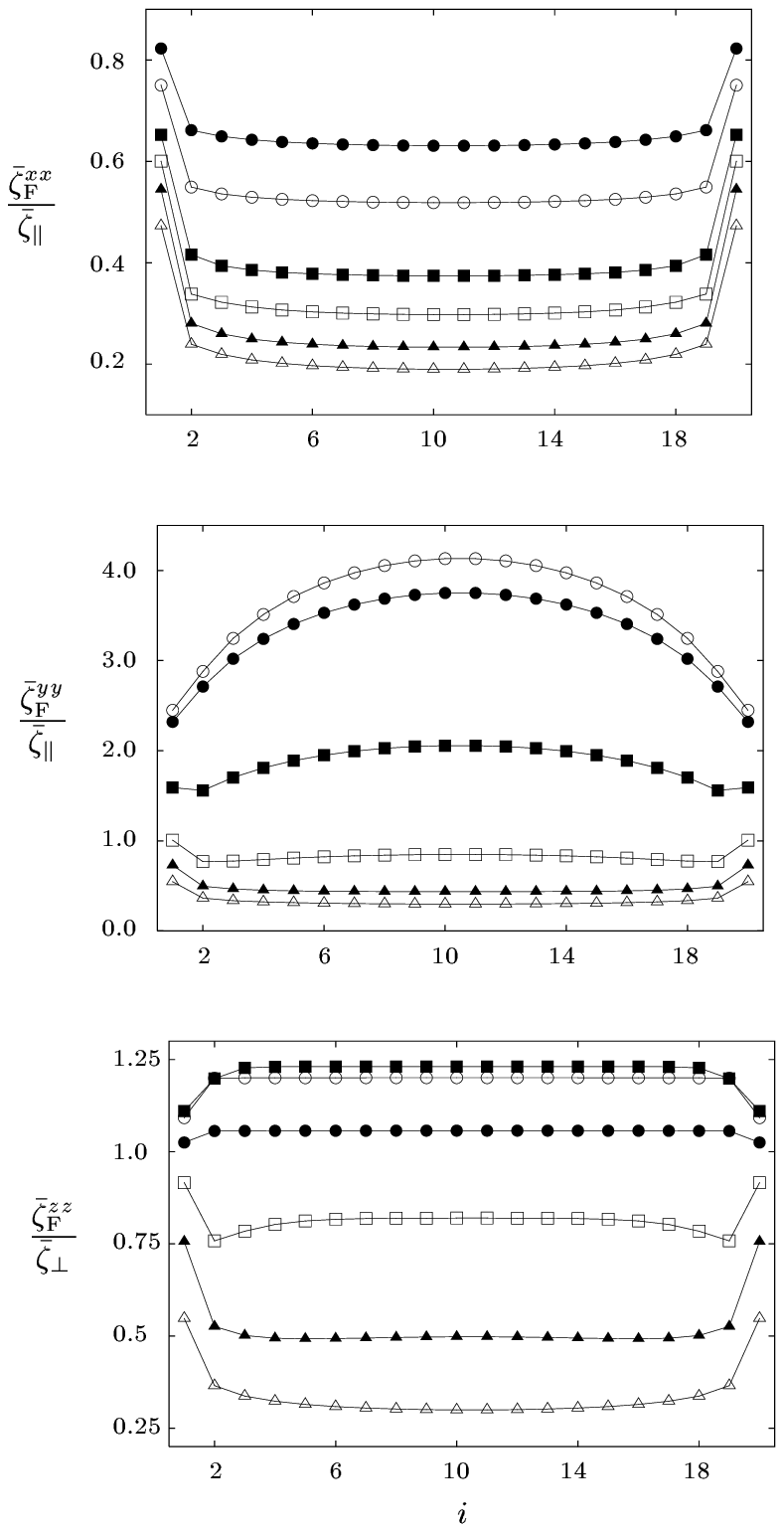}
\end{center}
\input{figtex/polymer-n20-force-cap}
\end{figure}

\begin{figure}
\begin{center}
\includegraphics{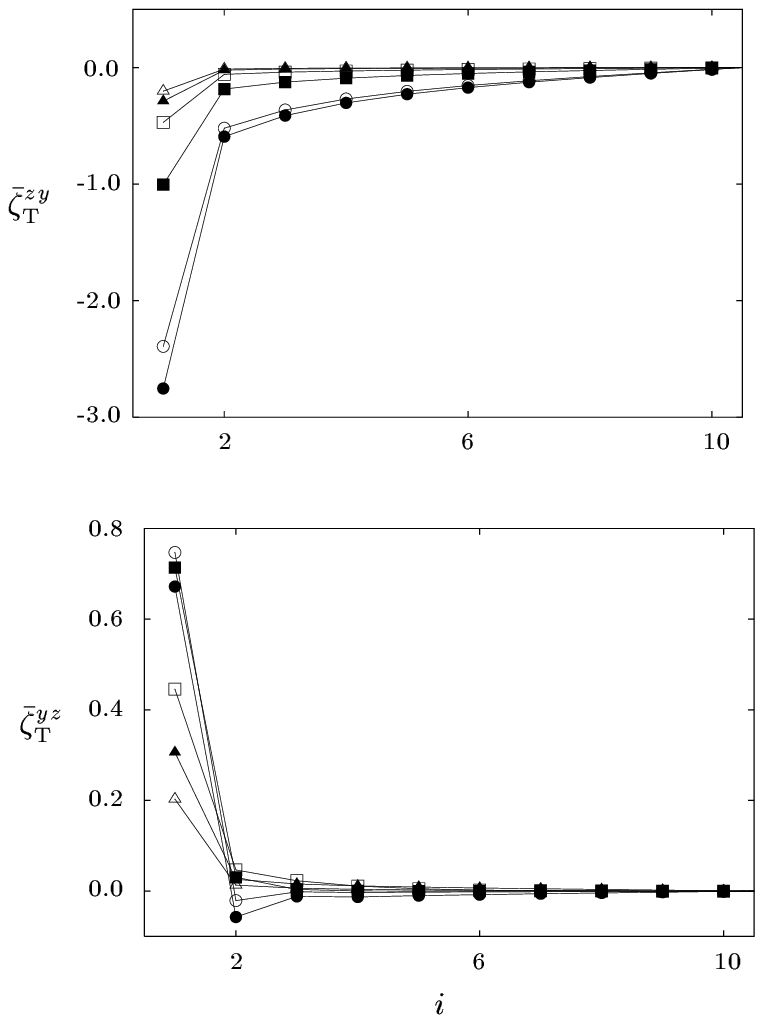}
\end{center}
\input{figtex/polymer-n20-torque-cap}
\end{figure}

\subsection{Multi-particle systems}
In figures \ref{rigid-body resistance}--\ref{torque on particles in
chain} we present some results for hydrodynamic resistance functions
of rigid linear arrays of $N$ touching spheres.  The spheres are
positioned on a line parallel to the axis $x$ at the center of the
space between the walls, i.e.,
\begin{equation}
\label{position of polymer}
h_i=\half H,\qquad i=1,\ldots,N.
\end{equation}

The diagonal components of the translational resistance
matrix of the array treated as a single rigid body, evaluated per one
sphere,
\begin{equation}
\label{rigid-body translational resistance}
\rigidBodyResistanceNorm^{\alpha\alpha}
   =(N\StokesResistance)^{-1}
    \sum_{i,j=1}^N
      \resistanceMatrixElement_{ij}^{\transl\transl\,\alpha\alpha},
\qquad\alpha=x,y,z,
\end{equation}
are plotted in figure \ref{rigid-body resistance} versus the number of
spheres $N$ in the chain.  The results for the longitudinal and
transverse components $\rigidBodyResistanceNorm^{xx}$ and
$\rigidBodyResistanceNorm^{yy}$ are shown normalized by the lateral
one-particle resistance coefficient $\bar\onePartResistanceLat$; the
vertical component $\rigidBodyResistanceNorm^{zz}$ is normalized by
$\bar\onePartResistanceTrans$.  The results indicate that for large
separations between the walls, when compared to the chain length, all three
components of the resistance matrix
$\rigidBodyResistanceNorm^{\alpha\alpha}$ decrease monotonically with
$N$, and behave as $1/\log N$ for $1\ll N\ll H/2a$.  We also find that
$\rigidBodyResistanceNorm^{yy}\simeq\rigidBodyResistanceNorm^{zz}
\simeq2\rigidBodyResistanceNorm^{xx}$ in this regime
\cite[][]{Blawzdziewicz-Wajnryb-Given-Hubbard:2005}.

For moderate and small values of the wall--to--wall distance $H$,
however, the behavior of each component
$\rigidBodyResistanceNorm^{\alpha\alpha}$ of the chain resistance
matrix is qualitatively different.  The longitudinal component
$\rigidBodyResistanceNorm^{xx}$ decreases monotonically with $N$,
which is similar to the behavior in the unbounded space, but the
variation is smaller.  The vertical component
$\rigidBodyResistanceNorm^{zz}$ initially increases with $N$, and then
saturates at a constant value that depends on the wall separation $H$.
In contrast, for small $H$, the transverse component
$\rigidBodyResistanceNorm^{yy}$ increases linearly with $N$ in the
range of chain lengths shown.  Additional numerical simulations for
chains with the length up to N=100 (not presented) indicate that the
resistance coefficients $\rigidBodyResistanceNorm^{yy}$ eventually
saturate for large $N$.  We note that the standard wall superposition
approximation entirely misses this behavior
\cite[][]{Bhattacharya-Blawzdziewicz-Wajnryb:2005a}.

A better insight into the mechanisms underlying the above-illustrated
qualitative features of the resistance matrix can be gained from the
set of more detailed results for a chain of the length $N=20$ plotted
in figures \ref{force on particles in chain} and \ref{torque on
particles in chain}.  In these figures, we show the resistance
coefficients
\begin{subequations}
\label{force and torque on individual spheres in chains}
\begin{equation}
\label{force on individual spheres in chain}
\individualSphereForceNorm{\alpha\alpha}(i) =\StokesResistance^{-1}
   \sum_{j=1}^N
   \resistanceMatrixElement_{ij}^{\transl\transl\,\alpha\alpha},
   \qquad \alpha=x,y,z,
\end{equation}
and
\begin{equation}
\label{torque on individual spheres in chains}
\individualSphereTorqueNorm{\beta\alpha}(i)=
      {\textstyle\frac{3}{2}}(a\StokesResistance)^{-1}
      \sum_{j=1}^N
         \resistanceMatrixElement_{ij}^{\rot\transl\,\beta\alpha},
\qquad \beta\alpha=zy,yz,
\end{equation}
\end{subequations}
representing the normalized applied force and torque
\refstepcounter{equation}
$$
\label{total force and torque on particle in chain}
\totForce_i=\hat\be_\alpha\individualSphereForceNorm{\alpha\alpha}(i),
\qquad
\totTorque_i=\hat\be_\beta\individualSphereTorqueNorm{\beta\alpha}(i)
\eqno{(\theequation{\mathit{a},\mathit{b}})}
$$ acting on particle $i$ in a chain moving in the direction $\alpha$
with a unit velocity.  By symmetry, the forces act only in the
direction of the chain motion, and the only nonzero torque
coefficients are those listed in equation \refeq{torque on individual
spheres in chains}.

According to the results shown in figure \ref{force on particles in
chain} for the motion in the $x$ direction, the forces acting on the
first and the last particle in the chain are larger than the forces
acting on the particles in the chain interior.  This behavior
is similar for chains in the unbounded and the wall-bounded regions.
The forces are smaller for long chains, because the particles
collectively drag the fluid in the direction of the chain velocity.
This mechanism is diminished, but not eliminated by the wall presence.

In an unbounded space, the force distribution in a chain moving in the
transverse direction $y$ is qualitatively similar to the distribution
for the longitudinal motion discussed above.  In the wall-bounded
region the results are, however, considerably different: the forces
near the center of the chain are much larger than the forces near the
chain ends. This behavior, clearly seen in figure \ref{force on
particles in chain} for $H/2a\lesssim2$, stems from the conservation of
the fluid volume.  The chain moving in the transverse direction acts
like a piston pushing fluid along the space between the walls, thus
producing a pressure-driven flow decaying on the lengthscale $l=2aN$.
The pressure increases linearly with the chain length until it is
large enough to push the fluid back through the gap between the walls
and the particles.  At this point, the pressure becomes independent of
N.  The pressure produced by this mechanism is responsible for the
large resistance coefficient $\rigidBodyResistanceNorm^{yy}$ of long
chains in transverse motion between closely spaced walls, as shown in
figure \ref{rigid-body resistance}.

For a chain moving in the direction $z$ (normal to the walls) in a
system with a small value of the wall--particle gap $\PWgap$, the
resistance coefficients $\rigidBodyResistanceNorm^{zz}$ are dominated
by the lubrication forces between the walls and the individual
particles.  The coefficients $\rigidBodyResistanceNorm^{zz}$ are the
smallest for the spheres at the chain ends, as seen in figure
\ref{force on particles in chain}, unlike for chains in the infinite
space.  This behavior stems from the presence of the geometrical
constraints---the resistance is smaller where there is more room for
the fluid to escape from the gaps between the walls and the particles.

The geometrical-parameter dependence of the torque acting on
individual spheres in a translating chain is less varied, as
illustrated in figure \ref{torque on particles in chain}.  In all
configurations considered, we find that the torque on the interior
spheres is much smaller than the torque at the chain ends. An
interesting feature is the sign change of the torque acting on
the particle $i=2$ for the coefficient
$\individualSphereTorqueNorm{yz}$.

\section{Conclusions}
\label{Conclusions}

Many-body hydrodynamic interactions in suspensions of spherical particles
confined between two parallel planar walls have been studied here
theoretically and numerically .  Our primary theoretical result is a set of
compact expressions for the multipolar matrix elements of the Green's integral
operator for Stokes flow in the space between the walls.  The matrix
elements are given in the form of lateral Fourier integrals of products of
several simple matrices.

Our expressions  have been used to develop an algorithm for evaluating
many-particle hydrodynamic friction and mobility matrices in a wall-bounded
suspension.  The algorithm involves solving a set of linear equations for the
multipolar moments of the force distributions induced on the particles.  The
resulting friction matrix is corrected for the lubrication forces by using a
superposition of particle-particle and particle-wall contributions. Our
algorithm yields highly accurate results---for example, the results presented
in this paper have been obtained with an accuracy better than $1\,\%$.  We
note that at each truncation of the force-multipole expansion the boundary
conditions at the walls are exactly satisfied.  This feature is essential for
obtaining a proper far-field behavior of the friction matrix, including the
strong backflow effects seen for rigid arrays of spheres.

Our numerical algorithm has been used to evaluate the hydrodynamic
resistance matrix for a single particle, a pair of particles, and a
system of many particles confined between two planar walls.  The
problem of hydrodynamic interactions in the two-wall geometry involves
several characteristic lengthscales: the particle radius $a$, the
wall--to--wall distance $H$, and the lateral distance between the
particles $\rho$.  For $\rho\ll H$ the interactions between particles
are similar to those in the infinite space.  For $\rho\approx H$ the
crossover occurs to a quasi-two-dimensional behavior in the regime
$\rho\gg H$.

In the quasi-two-dimensional domain the vertical components of the
mutual pair resistance matrix decay exponentially, and the lateral
components behave as $O(\rho^{-2})$.  Moreover, the sign of the
transverse component of the resistance matrix is opposite to the sign
of this component for a pair of particles in infinite space.  As
discussed here and in our recent paper
\cite[][]{Bhattacharya-Blawzdziewicz-Wajnryb:2005a}, this behavior can
be explained using the asymptotic Hele-Shaw (lubrication) form
of the far-field flow produced by a moving particle.

The crossover behavior is particularly pronounced for rigid arrays of spheres
arranged along a line parallel to the walls.  In the regime $a\ll l\ll H$,
where $l$ is the chain length, the hydrodynamic friction force per particle
decreases as $(\log l)^{-1}$ for large $l$, similar to the behavior in
the infinite space.  In contrast, for $l\gg H$ the longitudinal component of
the friction tensor (per particle) and the component normal to the walls tend
to constant values.  Moreover, for small particle--wall gaps, the transverse
component (normal to the chain but parallel to the walls) increases linearly
with the chain length before it saturates at a value that is much higher than
the corresponding value for the longitudinal motion.

As discussed in our recent paper
\cite[][]{Bhattacharya-Blawzdziewicz-Wajnryb:2005a}, the standard
wall-super\-position approximation is insufficient for many problems.
The resistance matrix in such an approximation is composed from two
single-wall contributions.  In particular, the superposition
approximation yields a wrong sign of the transverse component of the
mutual pair resistance matrix and a wrong far-field behavior of all
components of this matrix.  The approximation also fails to reproduce
the striking increase with the number of particles for the transverse
resistance coefficient of linear arrays of spheres.

The numerical efficiency of our method can be substantially improved by
combining our Cartesian representation of the wall contribution to the Green's
matrix with the asymptotic far-field expressions for this quantity.  The
asymptotic expressions which we have recently derived can be expressed in
terms of multipolar solutions of Laplace's equation for two-dimensional
pressure field corresponding to the lubrication flow in the space between the
walls.  These expressions do not involve Fourier integrals, and they can
relatively easily be implemented in numerical algorithms for periodic systems
and in accelerated PPPM or fast-multipole algorithms.

S.\,B.\ would like to acknowledge the support by NSF grant
CTS-0201131.  E.\,W.\ was supported by NASA grant NAG3-2704 and in
part by KBN grant No.\ 5T07C 035 22.  J.\,B.  was supported  by
NSF grant CTS-S0348175 and by Hellman Foundation.

\appendix

\section{Spherical basis}
\label{Spherical basis}

The spherical basis of Stokes flows $\sphericalBasisPM{lm\sigma}$ and
the reciprocal basis fields $\reciprocalSphericalBasisPM{lm\sigma}$
used in the present paper are normalized differently than the
corresponding basis fields $\bv^{\pm({\rm CFS})}_{lm\sigma}$ and
$\bw^{\pm({\rm CFS})}_{lm\sigma}$ introduced by
\cite{Cichocki-Felderhof-Schmitz:1988}.  The relations between these
sets of basis functions are as follows:
\begin{subequations}
\label{relation between BBW and CFS bases}
\begin{equation}
\label{relation between BBW and CFS v bases}
\sphericalBasisM{lm\sigma}(\br)
   =N_{l\sigma}^{-1}n_{lm}^{-1}\bv^{-({\rm CFS})}_{lm\sigma}(\br),
\qquad
\sphericalBasisP{lm\sigma}(\br)
   =N_{l\sigma}n_{lm}^{-1}\bv^{+({\rm CFS})}_{lm\sigma}(\br),
\end{equation}
\begin{equation}
\label{relation between BBW and CFS w bases}
\reciprocalSphericalBasisM{lm\sigma}(\br)
   =N_{l\sigma}n_{lm}r\bw^{-({\rm CFS})}_{lm\sigma}(\br),
\qquad
\reciprocalSphericalBasisP{lm\sigma}(\br)
   =N_{l\sigma}^{-1}n_{lm}r\bw^{+({\rm CFS})}_{lm\sigma}(\br),
\end{equation}
\end{subequations}
where
\begin{equation}
\label{change-of-normalization coefficients}
N_{l0}=1,\qquad N_{l1}=-{(l+1)^{-1}},\qquad N_{l2}=l[(l+1)(2l+1)(2l+3)]^{-1}, 
\end{equation}
and
\begin{equation}
\label{normalization coefficients}
n_{lm}=\left[\frac{4\pi}{2l+1} \frac{(l+m)!}{(l-m)!}\right]^{1/2}.
\end{equation} 

Below we list the explicit expressions for the angular coefficients
$\sphericalBasisCoefPM{lm\sigma}(\theta,\phi)$ for spherical basis
fields \refeq{spherical basis v +-} in our present normalization,
\begin{subequations}
\label{V-}
\begin{equation}
\label{V-0}
\sphericalBasisCoefM{lm0}=\frac{1}{(2l+1)^2}\left[
   \frac{l+1}{l(2l-1)}\alpha_l\bY_{l\,l-1\,m}
   -\frac{1}{2}\bY_{l\,l+1\,m}
\right],
\end{equation}
\begin{equation}
\label{V-1}
\sphericalBasisCoefM{lm1}=\frac{i}{l+1}\gamma_l\bY_{l\,l\,m},
\end{equation}
\begin{equation}
\label{V-2}
\sphericalBasisCoefM{lm2}=\beta_l\bY_{l\,l+1\,m},
\end{equation}
\end{subequations}
and
\begin{subequations}
\label{V+}
\begin{equation}
\label{V+0}
\sphericalBasisCoefP{lm0}=\alpha_l\bY_{l\,l-1\,m},
\end{equation}
\begin{equation}
\label{V+1}
\sphericalBasisCoefP{lm1}=\frac{\im}{l+1}\gamma_l\bY_{l\,l\,m},
\end{equation}
\begin{equation}
\label{V+2}
\sphericalBasisCoefP{lm2}=\frac{l}{2(2l+1)}\alpha_l\bY_{l\,l-1\,m}
   +\frac{l}{(l+1)(2l+1)(2l+3)}\beta_l\bY_{l\,l+1\,m},
\end{equation}
\end{subequations}
where
\begin{subequations}
\label{v-harmonics}
\begin{equation}
\label{vector harmonics A}
  {\bf Y}_{ll-1m}(\hr)=\alpha_l^{-1}
  r^{-l+1}\bnabla\left[r^l Y_{lm}(\hr)\right],
\end{equation}
\begin{equation}
\label{vector harmonics B}
  {\bf Y}_{ll+1m}(\hr)=\beta_l^{-1}
  r^{l+2}\bnabla\left[r^{-(l+1)} Y_{lm}(\hr)\right],
\end{equation}
\begin{equation}
\label{vector harmonics C}
  {\bf Y}_{llm}(\hr)=\gamma_l^{-1}\br\times\bnabla_{\mathrm{s}} Y_{lm}(\hr)
\end{equation}
\end{subequations}
are the  normalized
vector spherical harmonics, as defined by \cite{Edmonds:1960}.
Here
\begin{equation}
\label{scalar harmonics}
   Y_{lm}(\hr) =n_{lm}^{-1} (-1)^m P_l^m(\cos\theta)e^{{\rm i}m\varphi}
\end{equation}
are the normalized scalar spherical harmonics, and 
\begin{equation}
\label{norm.const.}
  \alpha_l=[l(2l+1)]^{1/2},\qquad
  \beta_l=[(l+1)(2l+1)]^{1/2},\qquad
  \gamma_l=-\im[l(l+1)]^{1/2}.
\end{equation}

\section{Transformation vectors 
$\frictionProjectionVector^\transl$ and
$\frictionProjectionVector^\rot$}
\label{Transformation vectors X}

The resistance matrix \refeq{resistance matrix} 
is obtained from the solution 
\begin{equation}
\label{solution of equation for force multipoles}
\inducedForceMultipole_i(lm)
   =\sum_{j=1}^N\sum_{l'm'}
      \GrandFriction_{ij}(lm\mid l'm')
   \bcdot
      \externalVelocityCoefficient_j(l'm')
\end{equation} 
of the force-multipole equation \refeq{induced force equations in
matrix notation} by projecting the generalized friction matrix
$\GrandFriction=\GrandMobility^{-1}$ onto the subspaces corresponding
to the rigid-body motion of the particle $j$ and the total force and
torque of the induced-force distribution on particle $i$.  As shown in
Ref.\ I, the projection can be expressed in the form
\begin{equation}
\label{Elements of physical friction matrix}
\resistanceMatrix^{AB}_{ij}
   =\sum_{lm\sigma}\sum_{l'm'\sigma'}
      \frictionProjectionVector(A\mid lm\sigma)
         \GrandFrictionElement_{ij}(lm\sigma\mid l'm'\sigma')
            \frictionProjectionVector(l'm'\sigma'\mid B),
\end{equation}
where $A,B=\transl,\rot$.   Here $\frictionProjectionVector(A\mid
lm\sigma)$ and $\frictionProjectionVector(l'm'\sigma'\mid B)$ are the
projection vectors defined by the equations
\begin{equation}
\label{friction projection matrix}
\frictionProjectionVector(\transl\mid lm\sigma)
   =\delta_{l1}\delta_{\sigma0}\tilde\frictionProjectionVector^\transl(m),
\qquad
\frictionProjectionVector(\rot\mid lm\sigma)
   =\delta_{l1}\delta_{\sigma1}\tilde\frictionProjectionVector^\rot(m),
\end{equation}
\begin{equation}
\label{friction projection vector translation}
\tilde\frictionProjectionVector^\transl(-1)=
({\textstyle\frac{2}{3}}\upi)^{1/2}
   \left[
      \begin{array}{c}
         1\\-\im\\0
      \end{array}
   \right],
\quad
\tilde\frictionProjectionVector^\transl(0)=
({\textstyle\frac{2}{3}}\upi)^{1/2}
   \left[
      \begin{array}{c}
         0\\0\\\sqrt{2}
      \end{array}
   \right],
\quad
\tilde\frictionProjectionVector^\transl(1)=
({\textstyle\frac{2}{3}}\upi)^{1/2}
   \left[
      \begin{array}{c}
         -1\\-\im\\0
      \end{array}
   \right],
\end{equation}
\begin{equation}
\label{relation between X r and x t}
\tilde\frictionProjectionVector^\rot(m)=-2\im 
\tilde\frictionProjectionVector^\transl(m),\qquad m=-1,0,1,
\end{equation}
and
\begin{equation}
\label{symmetry of friction projection matrices}
\frictionProjectionVector(lm\sigma\mid A)
   =\frictionProjectionVector^*(A\mid lm\sigma),
\qquad
   A=\transl,\rot.
\end{equation}

\section{Cartesian basis fields}
\label{Appendix on Cartesian basis fields}

The Fourier coefficients $\CartesianBasisCoefPM{\bk\sigma}(z)$ in
the expression \refeq{general form of Cartesian basis} for the
Cartesian basis fields are given by the expressions
\begin{subequations}
\label{Cartesian basis -}
\begin{eqnarray}
\label{Cartesian basis - 0}
\CartesianBasisCoefM{\bk0}(z)&=&(32\upi^2)^{-1/2}\,
\left[
   \im(1-2kz)\hat\bk+(1+2kz)\ez
\right]
   k^{-1/2},
\\\nonumber\\
\label{Cartesian basis - 1}
\CartesianBasisCoefM{\bk1}(z)&=&(8\upi^2)^{-1/2}\,
(
   \hat\bk\boldsymbol{\times}\ez
)
   k^{-1/2},
\\\nonumber\\
\label{Cartesian basis - 2}
\CartesianBasisCoefM{\bk2}(z)&=&(32\upi^2)^{-1/2}\,
(
   \im\hat\bk-\ez
)
   k^{-1/2},
\end{eqnarray}
\end{subequations}
and
\begin{subequations}
\label{Cartesian basis +}
\begin{eqnarray}
\label{Cartesian basis + 0}
\CartesianBasisCoefP{\bk0}(z)&=&(32\upi^2)^{-1/2}\,
(
   \im\hat\bk+\ez
)
   k^{-1/2},
\\\nonumber\\
\label{Cartesian basis + 1}
\CartesianBasisCoefP{\bk1}(z)&=&(8\upi^2)^{-1/2}\,
(
\hat\bk\boldsymbol{\times}\ez
)
   k^{-1/2},
\\\nonumber\\
\label{Cartesian basis + 2}
\CartesianBasisCoefP{\bk2}(z)&=&(32\upi^2)^{-1/2}\,
\left[
   \im(1+2kz)\hat\bk-(1-2kz)\ez
\right]
   k^{-1/2},
\end{eqnarray}
\end{subequations}
where $\hat\bk=\bk/k$.  The corresponding pressure fields are
\begin{equation}
\label{Cartesian pressure -}
   p^-_{\bk0}(\br)=(2\upi^2)^{-1/2}\eta\,
   k^{1/2}\e^{\im\bk\bcdot\brho-kz},\qquad
   p^+_{\bk2}(\br)=(2\upi^2)^{-1/2}\eta\,
   k^{1/2}\e^{\im\bk\bcdot\brho+kz},
\end{equation}
and
\begin{equation}
\label{Cartesian pressure +}
   p^-_{\bk1}(\br)=p^-_{\bk2}(\br)=
   p^+_{\bk0}(\br)=p^+_{\bk1}(\br)=0.
\end{equation}

\bibliographystyle{jfm} \bibliography{/home/jerzy/BIB/jbib}

\end{document}